\documentclass[11pt,a4paper]{article}

\usepackage{epsfig}
\usepackage{graphics}

\begin{document}

\title{CALICE Report to the DESY Physics Research Committee}


\author{\centering 
\LARGE\bf The CALICE Collaboration\footnote{A complete author-list can be found in Appendix 1.}
}


\maketitle

\vspace{-7cm}
\begin{flushright}
\end{flushright}
\vspace{5cm}

\begin{abstract}
We present an overview of the CALICE activities on calorimeter development for 
a future linear collider. We report on test beam analysis results, the 
status of prototype development and future plans. 
\end{abstract}

\newpage
\tableofcontents
\newpage

\section{Introduction}

The CALICE collaboration pursues the development of highly
granular calorimeters for a future
e$^+$e$^-$ linear collider, based on the particle flow approach
for optimal overall detector performance.

The collaboration consists of 57 institutes from 17 countries
in Africa, America, Asia and Europe and has 336 physicists and
engineers as members. Since the last review in 2007, the following
institutes joined the collaboration:
Tsinghua University, Beijing,
University of Bergen,
CERN,
Kyungpook National University, Daegu,
University of Kansas,
University of Louvain,
Max-Planck-Institute of Physics, Munich,
University of Rabat and Casablanca,
University of Tsukuba,
University of Wuppertal.

We follow different technological options for electromagnetic and hadronic
calorimeters in each case. Most of them are candidates for both particle 
flow
based ILC detector concepts, ILD and SiD, and for a detector at a 
multi-TeV
linear collider such as CLIC.
Our aim is to cover an as broad as possible range of options with
prototypes and test them in particle beams, thereby
maximizing the use of common infrastructures like mechanical devices,
ASIC architecture or DAQ systems. We work as much as possible in
a common software and analysis framework, in order to ease
combination and comparison of test beam data, and to achieve a
common understanding of strengths and weaknesses of the options
under consideration.

The major part of the effort is focused towards presenting realistic 
proposals
for the detector concept reports of the ILC technological design phase 1,
which is due in 2012. This is closely coordinated with the detector
concept groups. Realistic proposals must have the performance
established in test beams, and the designs must demonstrate the 
scalability
towards a full detector and provide estimates for dead regions
occupied by services and support, as well as a solid basis for costing.
Given the yet uncertain schedule and energy
range of the future collider, we also pursue developments which reach
a similar maturity only at a later stage.

The development of calorimeter prototypes generally proceeds in two steps.
Physics prototypes provide a proof-of-principle of the viability of a 
given
technology in terms of construction, operation and performance. In
addition they are used to collect large data sets for the study of 
hadronic
shower evolution with high granularity and the test of shower simulation
programs, and for the development of particle flow reconstruction 
algorithms
with real data. Technological prototypes address the issues of scaling,
integration and cost optimization.
Due to the different response of different active media to the components 
of
hadronic showers, the physics prototypes are needed for each active
material under consideration. Technical prototypes are needed for
each technology, but the effort can be kept reasonable by using common
building blocks, and by addressing large area and multi-layer issues 
separately without instrumenting a full volume. 

CALICE has completed data taking with physics prototypes of a silicon 
based
ECAL and scintillator based ECAL and HCAL; the test of a gaseous digital 
HCAL
is under preparation. Test beam data analysis is shifting emphasis from
establishing detector performance and understanding to shower model
validation and reconstruction development. Various technological
prototypes are at the stage of commissioning, establishing the
read-out chain and exposing detector elements to beam, in preparation of
beam tests with larger structures in the next years.

\section{Test beam analysis results}

\subsection{Test beam campaigns}

In the years 2008 and 2009 the CALICE collaboration
successfully commissioned and operated four calorimeter prototypes in the MTBF beam-line at FNAL 
(Chicago). The detector chain consisting of the silicon-tungsten electromagnetic calorimeter (Si-W ECAL), 
the scintillator-iron analogue hadronic calorimeter (AHCAL) and the tail catcher and muon tracker (TCMT) had 
already been previously tested at the SPS (CERN) in the years 2006-07.
The Si-W ECAL is a 30 layer sandwich structure with 1x1 cm$^2$ cell segmentation. The AHCAL is a 38 layer structure
with minimum cell size of 3x3 cm$^2$.
A new type of ECAL based on scintillator strips (4.5$\times$1 cm$^2$) and tungsten absorbers (Sc-W ECAL) replaced the Si-W ECAL 
during two test beam campaigns 
in September 2008 and May 2009.
In total four campaigns were supported by CALICE at FNAL, each of about 6 weeks. Fig.~\ref{fig:FNALTB} shows a step of the installation of the CALICE set-up at FNAL, next to a open active layer of the AHCAL prototype. 

In addition to these combined campaigns, several smaller beam test runs were 
performed for individual detectors and the results from these are described in
the relevant sections.

\begin{figure}
\centering \includegraphics[width=0.48\textwidth]{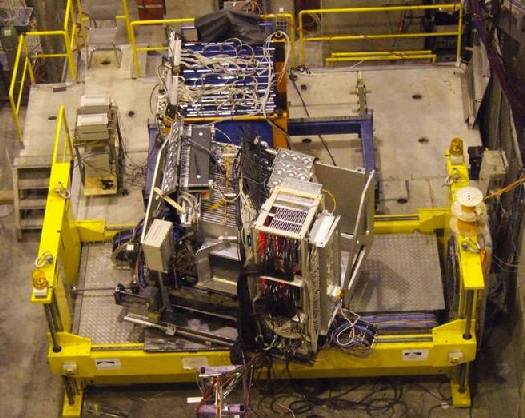}
\includegraphics[width=0.48\textwidth]{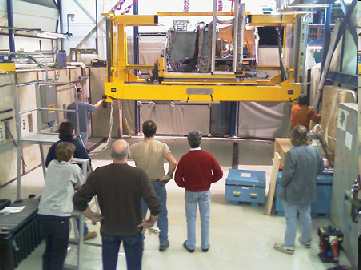}
\caption{\label{fig:FNALTB}\em (Left) CALICE test beam at CERN (Right) Installation of the CALICE movable stage in the MTBF area at FNAL. 
}
\end{figure}

The detectors were commissioned making large use of the charge injection and LED
monitoring systems.  In all four campaigns the commissioning phase took no
longer then one week.  The equalization of the cells' response was performed in
situ using dedicated calibration runs using a 32\,GeV muon beam.  The physics
program covered with both detector configurations included electron runs
in the energy range 1--30\,GeV, pions runs at 1--60\,GeV, 
and proton runs in the range 10--60\,GeV.
Particular interest for the physics program was to cover the region 1--6\,GeV not
covered at the CERN SPS.  In addition the Sc-W ECAL program included
special $\pi^0$ runs, in which the $\pi^0$ were generated by firing a 16--32\,GeV
$\pi^-$ beam at an additional target on the beam-line, immediately in front of
the detector.

Data were collected for both detector configurations under an
incident beam angle of 90 degrees to the detector planes (default configuration,
zero rotation) and in inclined positions of 80, 70 and 60 degrees, making use of
the movable stage support hosting the ECAL and HCAL detectors. In addition, $x-y$
scans were performed to study uniformity of response as a function of the beam
impact point on the detector surface.

A total of 87 million events were collected in the configuration 
including the Si-W ECAL, and 75 million events in the configuration 
including the Sc-W ECAL.

\subsection{Si-W ECAL performance}
\label{siwperf}
A description of the SiW ECAL physics prototype follows in section~\ref{SiWECAL_physproto}.
The Si-W ECAL was operated in the CERN beam tests in 2006-7, and in the 
first stage 
of the FNAL tests in 2008.  It was exposed to beams of muons, 
electrons and hadrons.
The muon data were used as the basis of detector calibration, so that 
recorded 
signals could be converted into minimum ionising particle (MIP) 
equivalents. 
The commissioning and calibration procedures have been described in some 
detail in~\cite{ECALcomm}.

\begin{figure}
\centering \includegraphics[width=0.48\textwidth]{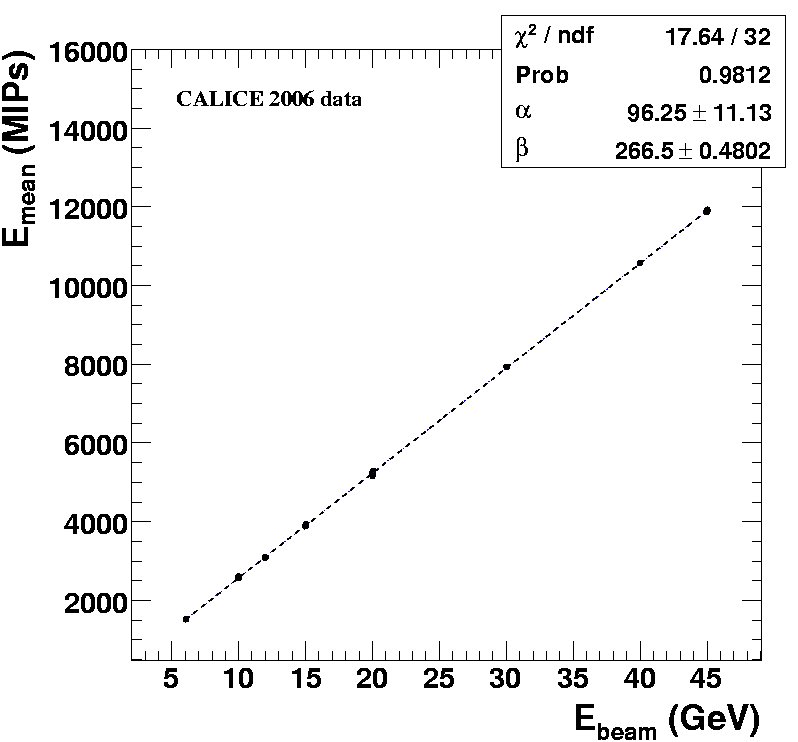}
\includegraphics[width=0.48\textwidth]{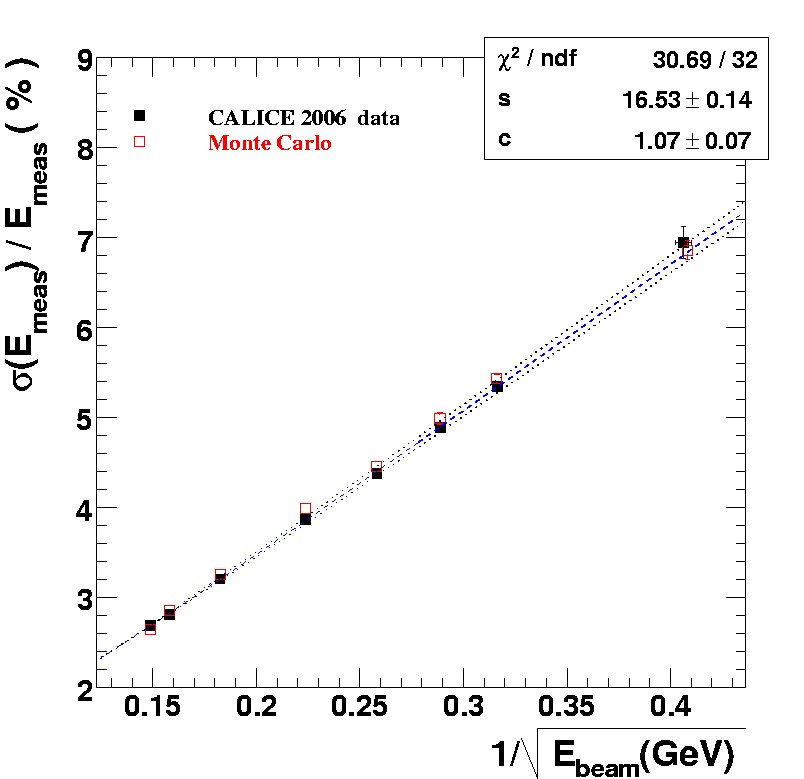}
\caption{\label{fig:EcalResp}\em (Left) ECAL response (in MIPs) as a 
function of beam energy.  A linear fit is superimposed. (Right) energy 
resolution as a function of $1/\sqrt{E}$, comparing data and simulation.  
A linear fit to data is shown.
}
\end{figure}

Some studies of the response of the ECAL to electron beams using the 2006 
data 
have been published in~\cite{ECALresp}.  Events are selected which lie 
well away 
from the edges of the test module and far from the inter-wafer gaps, in 
order to 
produce a clean sample for comparison with simulation.  In 
Fig.~\ref{fig:EcalResp} 
we show the measured response and energy resolution as a function of beam 
energy.  
The response is linear to better than 1\%, though with a small offset from 
the origin, 
which is largely predicted by the simulation.  The energy resolution is in 
agreement 
with that predicted for this configuration.  

\begin{figure}
\centering \includegraphics[width=0.85\textwidth]{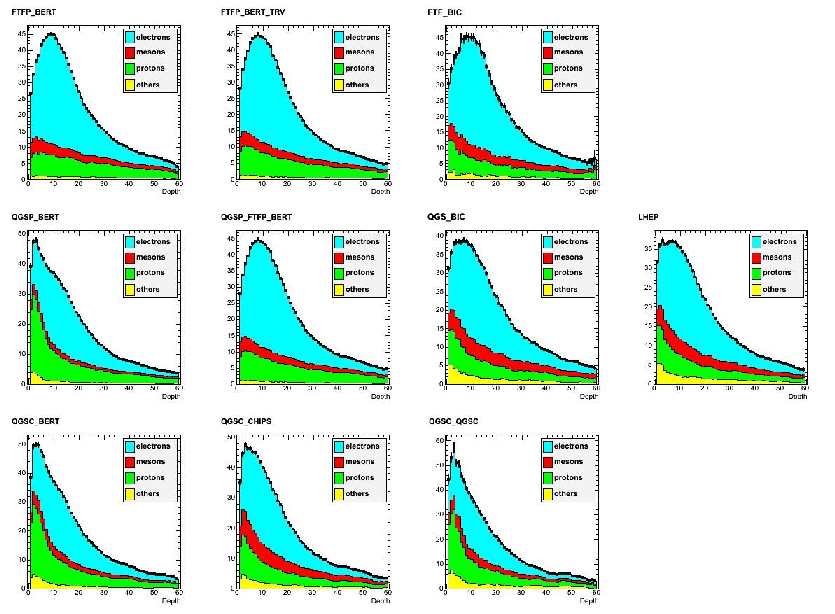}
\caption{\label{fig:EcalHadron}\em Monte Carlo simulations of the longitudinal energy profiles 
of 8\,GeV $\pi^-$ showers in the Si-W ECAL for ten different physics lists in GEANT4.
The energy is broken down into the contributions from e$^+$/e$^-$ (cyan), ``mesons'' (red -- $\pi^{\pm}$, 
K$^{\pm}$, $\mu^{\pm}$), protons (green) and ``others'' (yellow -- mainly nuclear fragments).  The ECAL granularity 
is clearly sufficient to offer sensitivity to these separate components, 
whose contributions differ between models, through their longitudinal    
development.}
\end{figure}

Many further studies using the ECAL data are in progress, for
example tests of uniformity using a transverse scan of the detector in
2007, measurements of transverse and longitudinal shower profiles, and
of position and angular resolution.  Several of these have been
presented in conference talks.  Studies are also under way of the ECAL
response to hadrons, which complement those using the HCAL. In
particular, the high granularity and short radiation length of the
ECAL permit interesting studies of the primary interaction in hadronic
showers to be made (see Fig.~\ref{fig:EcalHadron}).
 
\subsection{AHCAL performance}
\begin{figure}
\centering \includegraphics[width=0.48\textwidth]{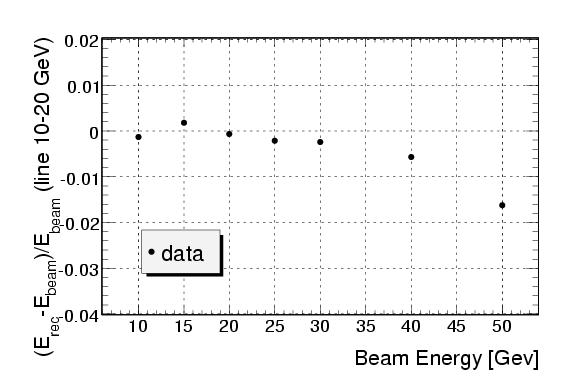}
\includegraphics[width=0.48\textwidth]{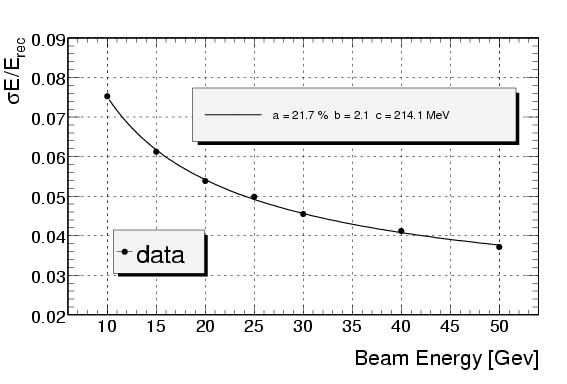}
\caption{\label{fig:HCALe}\em Left) Residuals of the reconstructed energy in
the AHCAL to a linear fit in the range
10-20 GeV. A linearity better than 2\% is obtained for electron energies up to 50
GeV. Right) AHCAL energy resolution for electrons.} \end{figure}

The completely instrumented AHCAL (described in section~\ref{sec:AHCAL}) 
was exposed to muon, electron and hadron 
beams in 2007-9, 
both with and without an ECAL in front.  Muons were used for calibration.  
An important 
test of our understanding of the calorimeter is to check the response to 
electrons 
with no ECAL in front of the AHCAL.  The energy density in electromagnetic showers is 
particularly 
large, so this is a good test of the important SiPM saturation 
corrections, and other effects~\cite{CAN-014}.  
In Fig.~\ref{fig:HCALe} we show the deviations from linearity of the electron response as a function of 
beam energy. Linearity better than 2\% is obtained for electron energies up to 50\,GeV.
The energy resolution for electrons is also shown.

\begin{figure}
\centering
\includegraphics[width=0.48\textwidth]{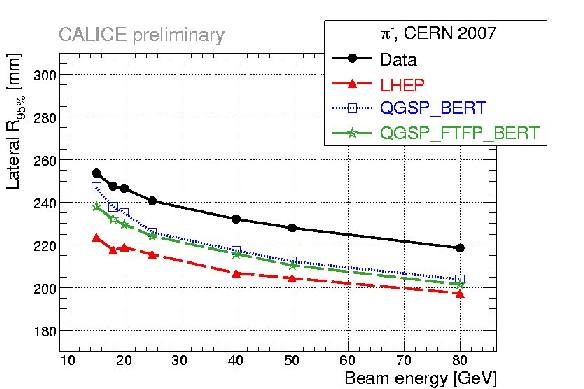}
\includegraphics[width=0.48\textwidth]{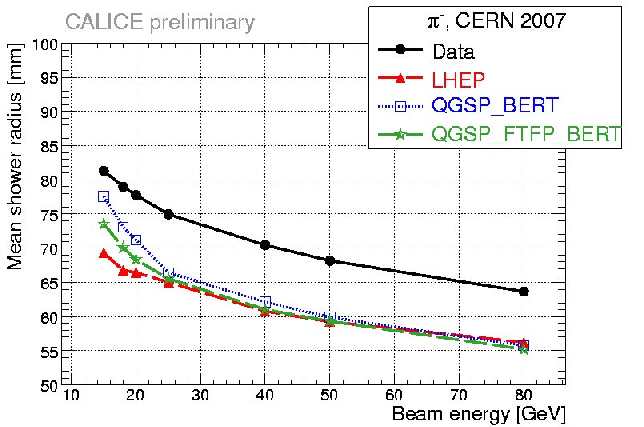}
\caption{\label{fig:HCALtrans}\em  Mean shower radius and the radius 
for containment of 95\% of the shower energy for pions in the AHCAL.  The 
data 
are compared with typical Monte Carlo simulations.
}
\end{figure}

The transverse and longitudinal energy profiles of hadronic showers are 
important characteristics, which will be expected to 
influence the performance of particle flow algorithms.
They can also be used to discriminate between Monte Carlo models.  
An example 
is shown in Fig.~\ref{fig:HCALtrans}, which is based on measurements of 
the
transverse shower profiles for pions, from which the mean shower radius 
and the radius 
for containment of 95\% of the shower energy are 
extracted~\cite{CAN-011e}.  
These are plotted 
against beam energy, and compared with the predictions of three of the 
many 
``physics lists'' available in GEANT4.  We note the need for caution in
drawing hasty conclusions about the merits of different models - the model 
which 
describes one observable best may perform less well for a different
variable, or at a different energy.  

\begin{figure}
\centering
\includegraphics[width=0.48\textwidth]{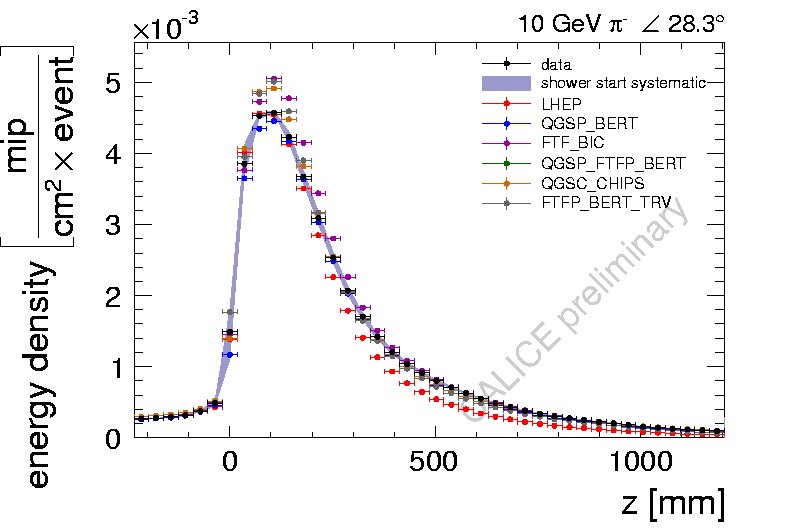}
\includegraphics[width=0.48\textwidth]{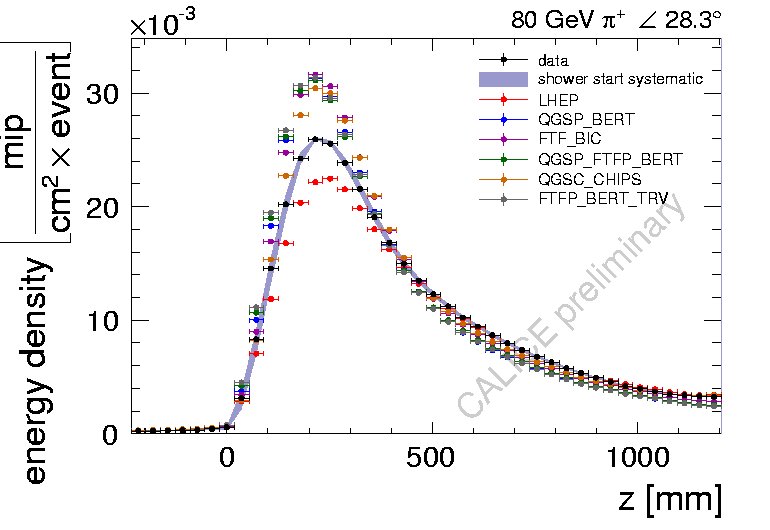}
\caption{\label{fig:HCALlong}\em  Longitudinal shower profiles
 for pions of 10\,GeV (left) and 80\,GeV (right) in the AHCAL.  The data 
are compared with typical Monte Carlo simulations.
}
\end{figure}

The longitudinal shower profile is important for particle flow, and 
can also be a useful discriminator between hadronic models. In order to
make a sensitive measurement, it is helpful to deconvolve the distribution 
of shower starting points from the form of the subsequent shower 
development.
We therefore developed an algorithm to identify the primary interaction 
point 
in the HCAL, and then measure the energy profile of the shower starting 
from this point.  Typical distributions~\cite{CAN-011d} 
are shown in Fig.~\ref{fig:HCALlong},
for pions of 10 and 80\,GeV, and compared with a few GEANT4 physics lists.  
None of the simulations is perfect at all energies, and the LHEP model 
seems 
particularly unsuccessful.  Work on studying the longitudinal shower 
development as a function of radius is also ongoing. 

\subsection{Combined performance}
\begin{figure}
\centering
\includegraphics[width=0.48\textwidth]{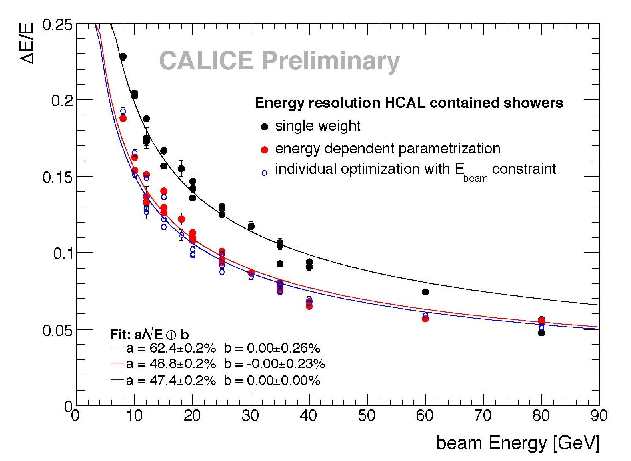}
\includegraphics[width=0.48\textwidth]{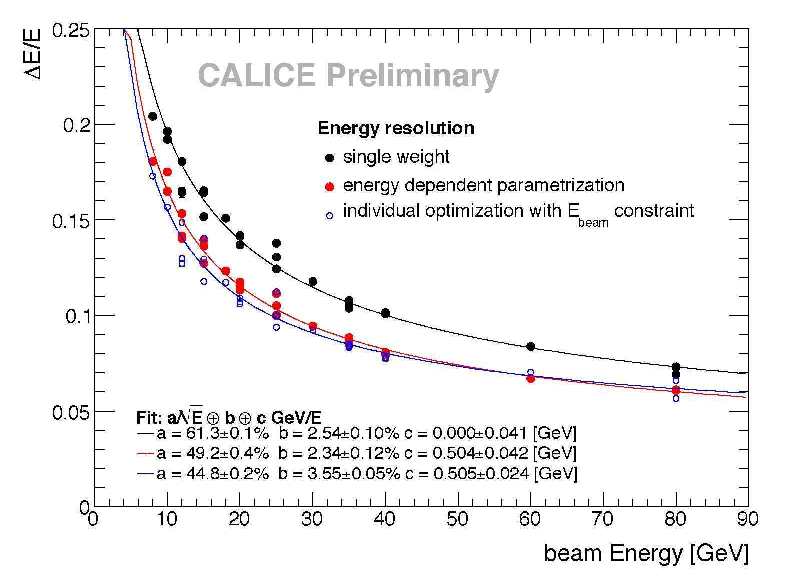}
\includegraphics[width=0.48\textwidth]{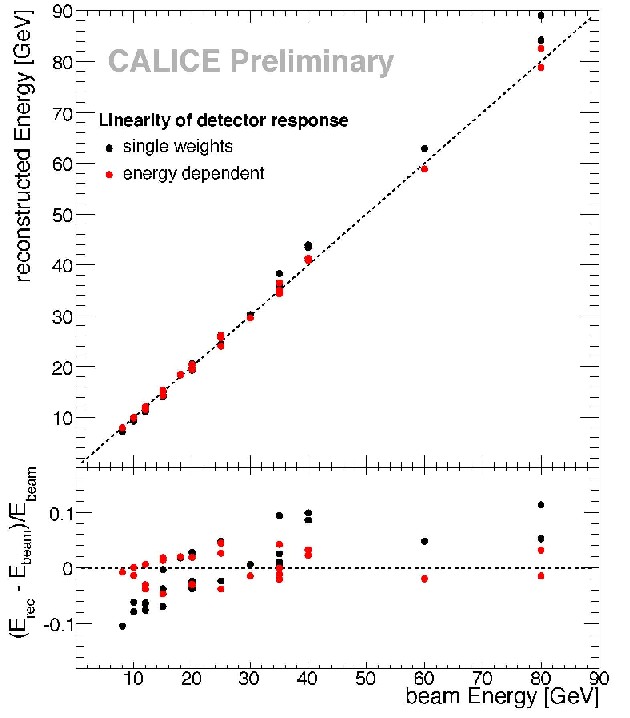}
\includegraphics[width=0.48\textwidth]{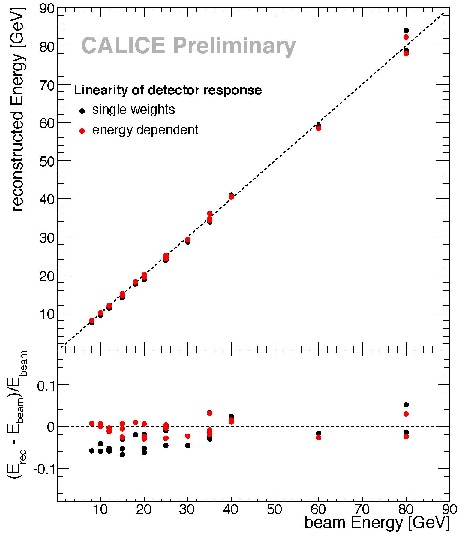}
\caption{\label{fig:HCALresol}\em  Energy resolution (above) 
and energy response (below) for pions as a function of pion energy, 
comparing the results with and without ``software compensation''.  The 
plots on the left are for showers confined in the AHCAL, while those on 
the right 
used the full ECAL/AHCAL/TCMT system.  }
\end{figure}

The CALICE calorimeters are non-compensating, i.e.\ they have different 
response 
to electrons and hadrons.  However, because of their 
high granularity, it is often possible to identify the various components 
within 
a shower, and then weight them appropriately.  One example of such 
``software 
compensation'' is illustrated in Fig.~\ref{fig:HCALresol}.  In this study, 
we exploited the fact that the electromagnetic components in a shower tend 
to have higher energy density.  Accordingly, cells are given different
weights depending on their energy.  We optimized these weights (in a 
parameterised form) so as to optimise the energy resolution.  As seen in 
Fig.~\ref{fig:HCALresol} (taken from~\cite{CAN-015}), 
the energy resolution for single pions can be improved from 
over $\sim$60\%$/\sqrt{E}$ to just below 50\%$/\sqrt{E}$.  A potential 
danger of this approach is that it might affect the linearity of the 
response, but in fact we find the linearity is slightly improved 
as well.  There is clearly scope for many future studies along these 
lines.      

\subsection{Sc-W ECAL performance}
\begin{figure}
\centering
\includegraphics[width=0.48\textwidth]{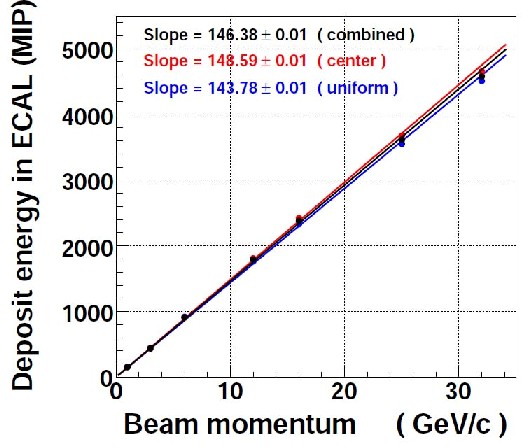}
\includegraphics[width=0.48\textwidth]{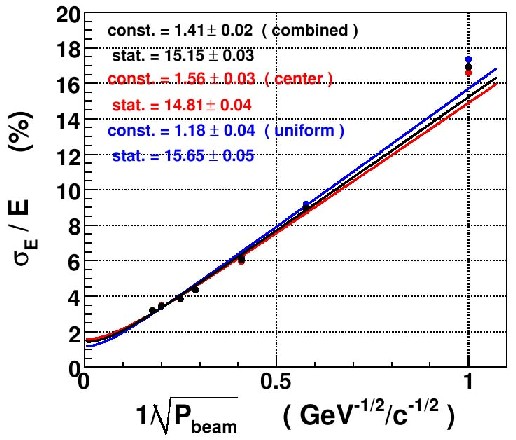}
\caption{\label{fig:ScECAL}\em  Energy response (left) and 
fractional resolution (right) as a function of beam energy for 
electrons incident on the Sc-ECAL.}
\end{figure}

A small prototype of the scintillator-tungsten ECAL was tested in 
stand alone mode at DESY in 2007. This was followed by full-scale beam
tests at FNAL 
in 2008 of a module of the same size as the Si-W prototype, with the
AHCAL and TCMT behind.  The scintillator was in the form of strips of 
size $\sim$$4\times 1$\,cm$^2$, read out by MPPCs via wavelength-shifting fibers.  
Some of the first results of these beam tests~\cite{CAN-016} are shown in 
Fig.~\ref{fig:ScECAL}.  At the present stage of understanding the data,
the response to electrons is linear to better than 5\%. An improved 
treatment of various corrections (temperature, for example)
can be expected to improve this.  The energy resolution is also 
shown, and is rather comparable to that of the Si-W ECAL. 

\section{Silicon-tungsten ECAL: Si-W ECAL}
\label{SiWECAL_physproto}
\subsection{Physics prototype}

A small detector prototype, the ``physics prototype'', was developed
in order to study the physics performance of a sampling electromagnetic calorimeter
suitable for use in detector at a future Linear Collider. 
The detector was designed to satisfy the requirements
of the Particle Flow approach to event reconstruction, in particular
a compact size, small Moli\`ere radius, and high granularity.
A 30--layer sampling calorimeter design was chosen, with three
different sampling fractions, finest in the first layers and
coarser in the latter layers. Tungsten was used as absorber material
due to its small Moli\`ere radius and large interaction to radiation
length ($\lambda_I / X_0$) ratio.
The active layers were made of 525\,$\mu m$ thick silicon sensors,
segmented into cells of $1\times 1$\,cm$^2$, giving the high
granularity required for the particle flow approach.

\begin{figure}
\centering
\includegraphics[width=0.45\textwidth]{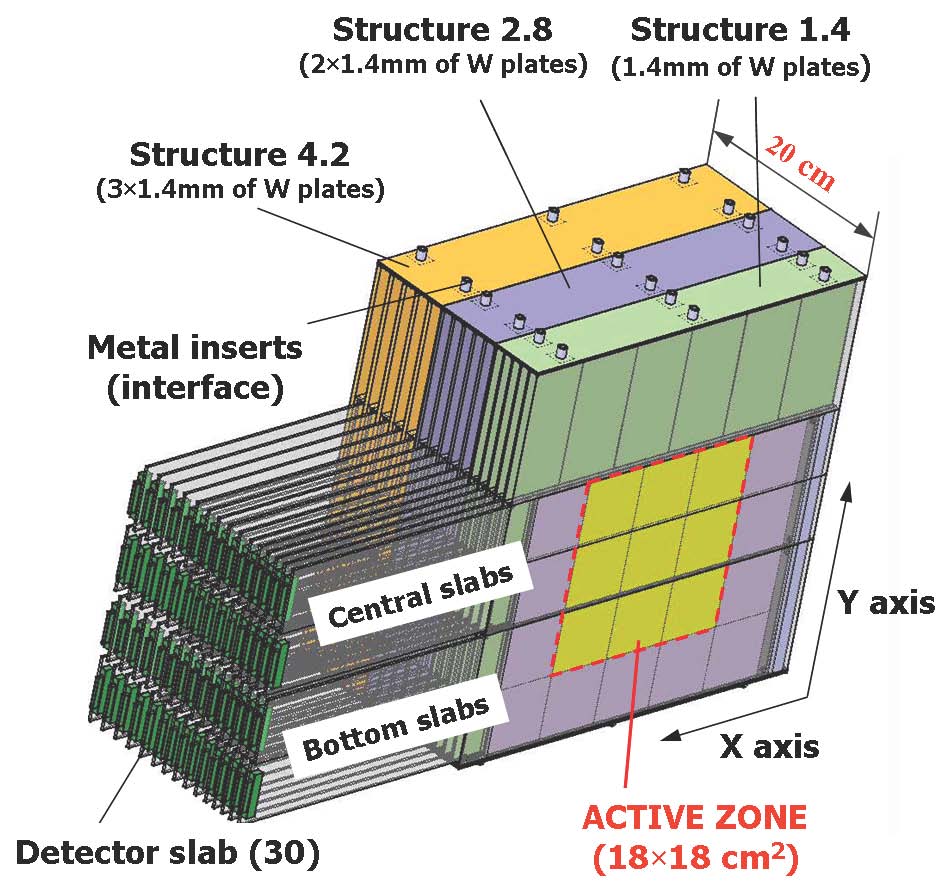}
\includegraphics[width=0.45\textwidth]{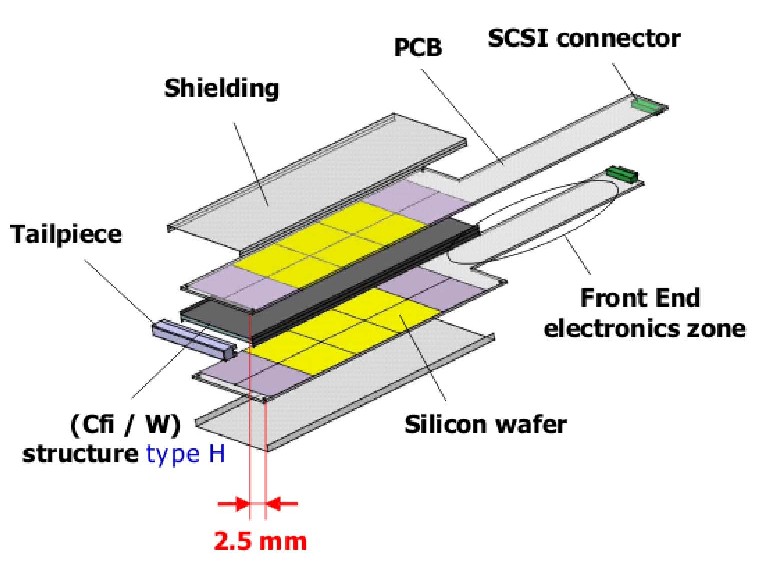} \\
\includegraphics[width=0.45\textwidth]{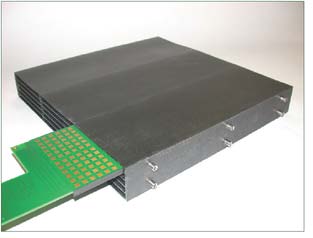}
\includegraphics[width=0.25\textwidth]{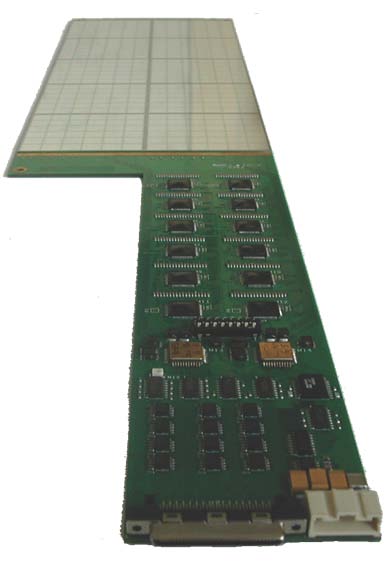}
\caption{\label{fig:PhysProto}\em Diagrams (upper) and photos (lower) showing 
the alveolar support structure (left) and detector slab (right).}
\end{figure}

These silicon detectors were mounted on printed circuit boards which 
channelled the signals to the front-end readout chip, mounted on the same
PCB outside the detector volume. Two PCBs were mounted on 
each side of an ``H''-shaped detector slab which also incorporated a layer 
of tungsten.
These detector slabs were then inserted into an alveolar supporting 
structure
composed of a carbon fiber-epoxy composite material, also incorporating
the other half of the tungsten absorber layers.

\subsection{Operational experience with physics prototype}

This ECAL prototype was exposed to test beams in 2006-07 (at CERN)
and 2008 (at FNAL).
Several hundred million events were collected in total, 
comprising a mixture of calibration events and beam
events with different particle species and momenta. 
Beams of electrons, positrons, muons,
pions and protons were used at a range of momenta between 1 and 180\,GeV/c.
The detector ran stably over this three year operation period, in terms of 
calibration, energy response and linearity. 
No adverse ageing effects were observed.

In the next generation ECAL prototype (see below) the front end 
electronics
will be placed inside the detector volume. A concern is that the
electronics chips might produce spurious signals when they lie in the
centre of a dense electromagnetic shower. To test this, a special
detector slab was prepared, consisting of four readout chips, 
without any silicon wafers. 
This layer was placed into the physics prototype structure at the depth 
of the
electromagnetic shower maximum, 
and exposed to high energy electrons. An analysis of 
the
resulting data has shown that there are no effects of fake signals due to 
this effect.

\subsection{Physics prototype detector performance}

The electron data collected have been used to measure the detector 
performance, as described in Sec.~\ref{siwperf}.


Several inconvenient features were encountered in the operation of the detector.
Some chips showed unstable pedestal values during the running period,
requiring offline corrections to be applied to the data.  This feature was understood, 
and will be remedied in future designs. 

\begin{figure}
\centering
\includegraphics[width=0.45\textwidth]{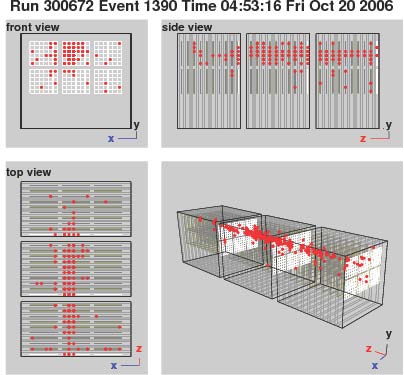}
\includegraphics[width=0.45\textwidth]{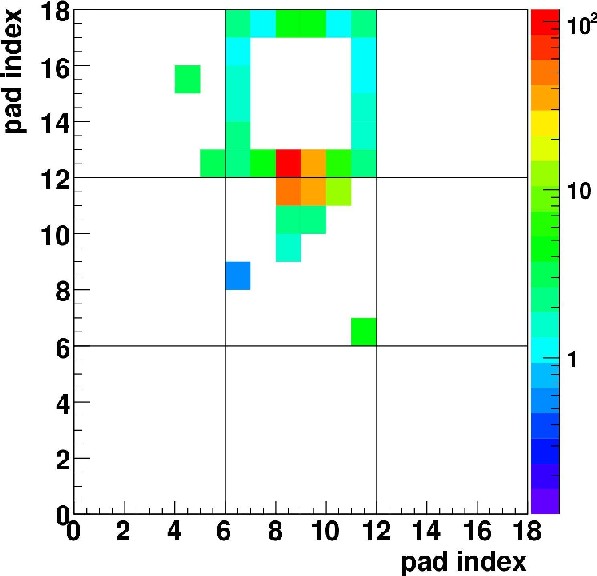}
\caption{\label{fig:eventDisplay}\em Event display of normal and ``square'' 
event.}
\end{figure}

An interesting problem was observed when an electron deposited energy
in the guard ring (a structure at the edge of the silicon sensor which
protects against high voltage breakdown). The deposited charge propagates 
around this structure, which, via the coupling between the guard ring
and adjacent cells, produced events containing distinctive square shapes,
with signals seen in all the cells at the sensor edge (see Fig.~\ref{fig:eventDisplay}). 
This feature of the sensor has prompted R\&D to improve its design.


\subsection{Conceptual issues for the technical prototype}

\begin{figure}
\centering
\includegraphics[width=0.45\textwidth]{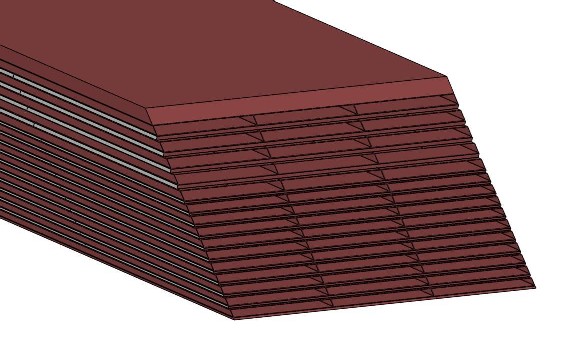}
\includegraphics[width=0.45\textwidth]{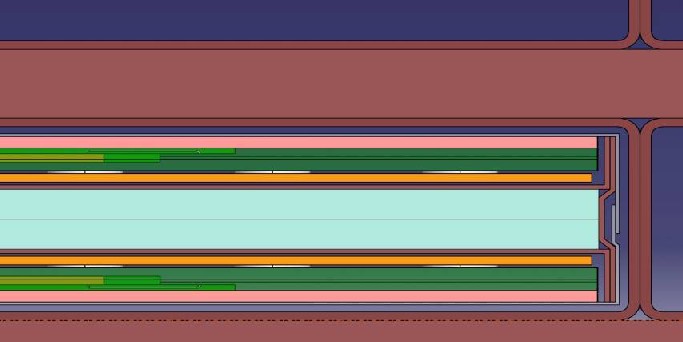} \\
\caption{\label{fig:eudet}\em Design of the Si-W ECAL technical prototype.}
\end{figure}

The next stage towards the realisation of a final detector for a Linear 
Collider detector is the production of a technical prototype. This
is conceived as a slightly smaller scale version of a single module
of the final ECAL detector. It has the same shape as the ILD ECAL module
design, the same number of layers, but somewhat smaller transverse 
dimensions (see Fig.~\ref{fig:eudet}).

This next generation of silicon detectors will have a granularity
four times higher than in the physics prototype, with a cell size of 
$5\times5$\,mm$^2$.
To minimise the effect of dead areas at the edge of silicon sensors, 
larger
$9\times9$\,cm$^2$ sensors have been developed. The design of the 
sensors'
guard rings is under investigation to minimise the propagation of signals
along the ring and the appearance of the so-called ``square events'', 
mentioned above.

In contrast to the physics prototype, where the front end electronics
was placed outside the detector volume, the technical prototype
will have the electronics directly embedded in the PCBs which support the
silicon detectors. This requires the design of ambitiously thin and complex
PCBs, and studies of the bonding and encapsulation of unpackaged chips.

The front end chip must consume very little power to prevent massive 
cooling requirements. They should in particular make use of the ILC bunch 
structure, where beam is delivered to the detectors only $\sim 1\%$ of the
time. A power-pulsed design of the electronics will allow the chips to 
be powered for only the time when beam is delivered (plus some readout 
time), 
and will be powered down for the remaining $\sim 99\%$ of the time.

The zero-suppressed, digitised signals from the front-end electronics 
are then passed to the common CALICE DAQ system (see Sec.~\ref{DAQsection}).

\subsection{Status and plans for the technical prototype}

\begin{figure}
\centering
\includegraphics[width=0.45\textwidth]{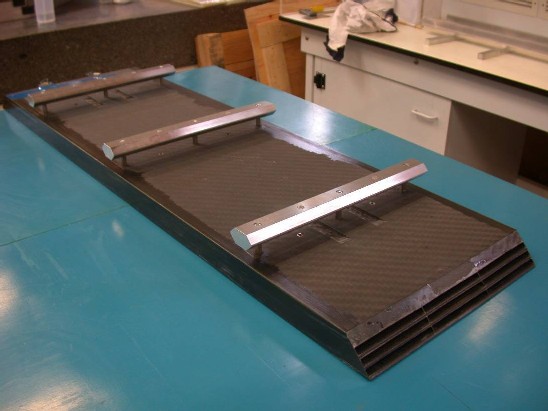}
\caption{\label{fig:demonstrator}\em Photo of mechanical demonstrator module.}
\end{figure}

To study the fabrication of the tungsten-composite mechanical alveolar
structure, an intermediate step, a three-layer ``demonstrator'' module has been 
constructed (Fig.~\ref{fig:demonstrator}).
This allowed the study of composite layer manufacture, cutting and
finishing, as well as the assembly of the various layers into a
final structure, together with tungsten plates and the thick composite 
back-
and front-plates which provide the rigidity of the module.
The manufacture of this module was a success, and the measured dimensions
of the module satisfy the mechanical requirements.
The ECAL will be fixed to the inner surface of the HCAL via a system
of rails on the outer face of the ECAL modules. The demonstrator module
incorporated such rails.

Work has recently started on the manufacture of the larger alveolar layers
for the technical prototype. A first layer has been constructed using
new moulds to reflect the larger size of the detector slabs for this 
prototype.

Research and development of the silicon sensors is in progress. 
Forty $9\times9$\,cm$^2$ wafers with $5\times5$\,mm$^2$ pixels have 
been supplied by Hamamatsu. 
A number of these sensors have been tested and show satisfactory 
performance.
An ongoing collaboration together with OnSemi and the Institute of Physics 
(Prague)
is studying the design of the guard rings to minimise the occurrence of 
``square events''.
A number of small test sensors have been produced with different guard 
ring structures.
Tests show that these designs do indeed reduce the propagation of signals 
around the
wafer edge.

In the medium term, links will be forged with further industrial partners. 
This will introduce
an element of competition to the sensor production, which should help to 
reduce
the sensor cost, which is at present uncomfortably high when a complete 
ILC detector ECAL is considered.
The EUDET ECAL module will be instrumented with sensors from a number of 
different producers.

The construction of the detector slabs is under study. A long string of
up to $\sim$ 10 Active Sensor Units (ASU, the PCB supporting the silicon
detectors)
must be connected together. This connection is both mechanical and
electrical - to supply power and send and receive data to the front-end 
chips.
There are strong constraints on the available space for these connections,
so a very thin, mechanically and electrically robust system is under 
design.
A dedicated assembly bench has been produced, which allows safe and well 
controlled
manipulation of the delicate ASU elements. Studies of ASU connection are 
underway.

The front-end chips are being designed. The EUDET module will be equipped 
with
SKIROC2 chips, see Sec.~\ref{sec:FEE}.
Since these are not yet available, 
in the meantime tests will be carried out
using the SPIROC2 chip. This chip can be operated in a ``SKIROC mode'', in 
which
its behavior is rather close to that envisaged for the SKIROC2 chip; 
however it has a smaller
number of channels than the SKIROC2 chip, allowing only a subset of 
silicon cells to be read out. 
The operation of this chip in SKIROC mode has been tested on an 
electronics test-bench, and shows the
expected characteristics.

The design of the PCB is under study. The PCBs for the technical 
prototype
will, in contrast to the PCBs used in the previous prototype, will hold 
the
front-end chips inside the detector volume. The space available for the 
PCB
is limited, to avoid degrading the detector performance; it must have a 
height
of not larger than 1.2mm. This constraint places particular emphasis on 
the
integration of the chips onto the PCBs and on how they are bonded.
A functional version of the board with relaxed requirements on the 
thickness
has been manufactured, and is being tested. Further prototypes are being 
designed
and submitted to manufacturers.

A cosmic test-bench is under preparation, which will allow the whole chain 
of 
sensors, PCB, front-end electronics and DAQ system to be tested and 
debugged.

The removal of heat produced in the front-end electronics from the 
detector structure
is essential for the integration of the ECAL into the general detector.
At present a 500\,$\mu$m sheet of copper extends along the length of the 
detector
slab, acting as a thermal drain. The heat is extracted from this copper 
drain at the
end of each module by a water-cooled system. Several possible designs for 
this
system are under consideration, and a number of prototypes have been 
built.

We plan to instrument a $18\times 18$\,cm$^2$ tower of the next prototype 
(consisting of 30 layers)
with sensors. One long detector slab will be prepared, to allow the 
testing of signal
propagation along the entire $\sim$1.5\,m length of a detector slab. 
The remainder 
will be composed of short detector slabs, each holding $2\times 2$ silicon 
sensors per layer.
The detector will be gradually instrumented as silicon wafers become 
available.
A series of cosmic and beam tests will allow tests of the whole detector 
system, even with a partial 
complement of silicon elements. Combined tests with other detectors (HCAL, 
tracking, muon)
are foreseen as part of the CALICE program.

\section{Scintillator-tungsten ECAL: Sc-W ECAL}
The scintillator ECAL group which consists of 4 universities from Asia,
Kobe, Kyungpook, Shinshu and Tskububa, is developing an
EM calorimeter with tungsten absorber and scintillator strips according
to the idea of PFA. In order to satisfy the requirements of PFA, the ECAL is expected to have fine segmentation of the order of a cm.
To achieve this with scintillator we arrange  plastic
scintillator strips with 1cm width orthogonally in successive layers. 
A strip is read out by a small silicon photo-sensor called MPPC 
via a wavelength shifting fiber inserted in a hole in the scintillator strip.

An example of integration of the system as a calorimeter, we produced a small
prototype which was tested at the DESY test beam in 2007. Then we have 
constructed a four-times bigger physics prototype and tested it at FNAL in years '08-'09.

\subsection{Operational experience with the physics prototype}

\begin{figure}
\centering
    \includegraphics[angle=0,width=0.7\textwidth]{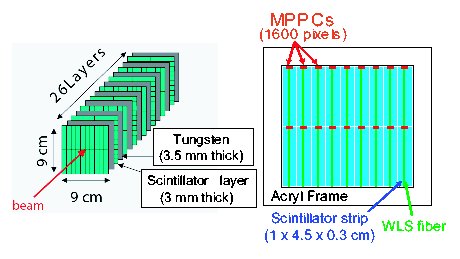}
\caption{\label{fig:tohru1}\em Sketch of the Sci-W ECAL structure.}
\end{figure}

We have constructed a 26 layer Sci-W ECAL with 18 strips/layer. The
detector has a sandwich structure of 3.5 mm tungsten and 3 mm scintillator. 
The absorber layer is composed of 88\% tungsten, 12\% cobalt and
about 0.5\% carbon, and has Moliere radius of 10 mm.
The scintillator layer has
9$\times$2 scintillator strips of size 4.5$\times$1 cm$^2$. 
In successive scintillator layers,
the strips are alternately aligned vertically (X layers) and horizontally (Y
layers).
The scintillation
light is read out by MPPCs (produced by Hamamatsu Photonics Co.). 
The MPPC is one of the novel pixelated Geiger Mode
Avalanche photo diode, such as SiPM. A MPPC consists of a matrix of micro APD
pixels of 25 $\mu m^2$. The total number of pixels in a device is 40$\times$40=1600 pixels
in 1 mm$^2 $. The structure of the Sci-W ECAL test module is shown in Figure~\ref{fig:tohru1}. 

\begin{figure}
\centering
    \includegraphics[angle=0,width=0.7\textwidth]{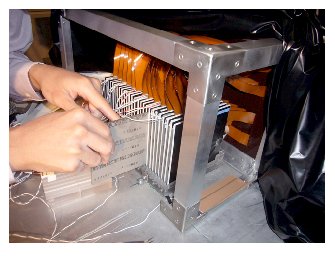}
\caption{\label{fig:tohru2}\em Small size Sci-W ECAL prototype installed at the DESY test beam.}
\end{figure}

The MPPCs are mounted
in a hole in the end of each scintillator strip and soldered to a flat cable. 
Signals from the MPPCs are fed into
the AHCAL baseboard through the flat cable. Figure~\ref{fig:tohru2} shows a photograph of
the Sci-W ECAL test module mounted in the beam line. The size of the entire module
is about 9$\times$9$\times$20 cm$^3$ and the total number of readout channels is 468.

The module was tested at DESY with positrons at energies between 1 and 6 GeV.
We use a combination of trigger and veto counters to trigger events. 
The trigger counters consist of a 3$\times$3$\times$1 cm scintillator block read out by two photo-multipliers. Two trigger counters and
one veto counter were installed in the beamline upstream of the ECAL module
and used to select 1 MIP events. One veto counter was located downstream of
the ECAL and used to select positrons which do not shower during their passage
through the ECAL when active layers were exposed outside the absorber structure. Such MIP events are used to calibrate the detector response.

Each strip was calibrated by these MIP events.
Four layers of drift chambers were placed in the beamline
to perform precise particle tracking. Each drift chamber layer has both x and y
wires. The ECAL test module was mounted on a movable stage, allowing it to
be moved with 0.1 mm accuracy in the plane transverse to the beamline.

\begin{figure}
\centering
    \includegraphics[angle=0,width=0.4\textwidth]{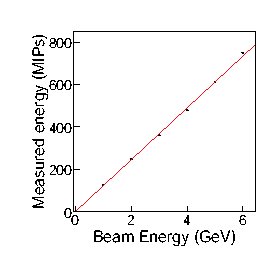}
    \includegraphics[angle=0,width=0.5\textwidth]{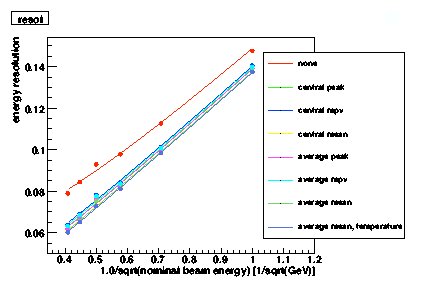}
\caption{\label{fig:tohru5}\em Linearity (left) and energy resolution (right) of the small prototype of Sci-W ECAL tested at the DESY test beam.}
\end{figure}

The energy response of the calorimeter is
shown in figure~\ref{fig:tohru5} (left) which indicates good linearity.
The resolution is also derived from these measurements as shown in Figure~\ref{fig:tohru5} (right).
The structure of the Sci-W ECAL test module is shown in Figure~\ref{fig:tohru1}. 

\begin{figure}
\centering
    \includegraphics[angle=0,width=0.7\textwidth]{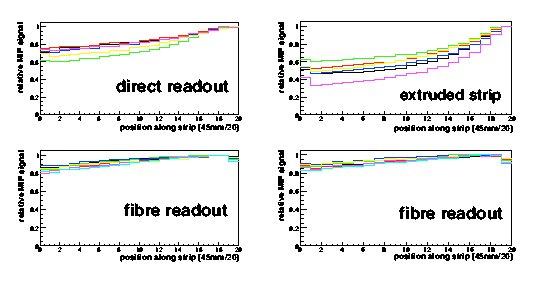}
\caption{\label{fig:tohru7}\em Measured non-uniformity of various scintillator strips used in the Sci-W ECAL.}
\end{figure}

The red dots and red line are corresponding to the extruded scintillator which was introduced at the first time for this test. The difference is expected from a
simulation to be due to the non-uniformity of strips along the wave length shifting fiber , which is measured as shown in figure~\ref{fig:tohru7}.

In this test beam experiment we have employed three different types of scintillator from two different
manufactures. The direct and fiber readout scintillator are made by Kuraray, 
and the extruded strips are produced by Misung CO. in Korea. The non-uniformity
is clearly larger for extruded ones. Since the cost of the scintillator for the
extruded is quite low, we decided to improve the performance for the
extruded in Korea for the next beam test at FNAL.

\begin{figure}
\centering
    \includegraphics[angle=0,width=0.45\textwidth]{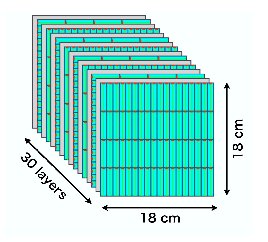}
    \includegraphics[angle=0,width=0.45\textwidth]{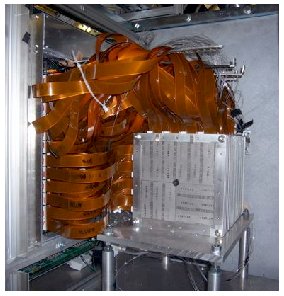}
\caption{\label{fig:tohru8}\em Sketch (left) and picture (right) of the large Sci-W ECAL prototype installed in the FNAL test beam area.}
\end{figure}

Subsequently, we have built the Sci-W ECAL with MPPC and extruded scintillator
with 4 times bigger volume, which is shown in figure~\ref{fig:tohru8}.

The second Sci-W ECAL prototype incorporates the improvements suggested by the 
results of the first test beam experience, 
although keeping the readout scheme and electronics. It is 18$\times$18 cm$^2$ in
lateral size and composed of 30 layers. The number of radiation lengths of this
calorimeter is 21 X$_0$, although the same tungsten plates were employed. A
scintillator layer consists of 4 x 18 =72 strips of 1$\times$4.5 cm$^2$ with 3 mm
thickness. Even layers have X-strip and odd layers have Y-strips with WLSF read
out covered by a good reflector film of KIMOTO CO.. This prototype has been tested 
during two
comprehensive beam test periods at FNAL, in Sep. 2008 and
May 2009. The prototype was operated together with the 
AHCAL and the TCMT to evaluate the combined performance of calorimeters within
the framework of CALICE. 
Electron and charged pion beams with an energy between 1 GeV and 32
GeV were used. 
Beams were injected into the central region and
uniform region, to confirm the improvement in uniformity of scintillator
response. The detector was calibrated using a 32 GeV muon beam. 

Using muons as minimum ionizing particles the MIP calibration constants for 
2160 channels is 160 ADC counts.
It is a factor 10 larger than the corresponding standard
deviation of the pedestal which is 15 ADC counts. The statistic uncertainty of the
MIP response on each channel is less than 0.6\% even in the worst channels.
There are five dead channels due to broken photo-sensors or too high levels of
background noise. The contribution of the dead channels to the energy
measurement is negligible.

\begin{figure}
\centering
    \includegraphics[angle=0,width=0.6\textwidth]{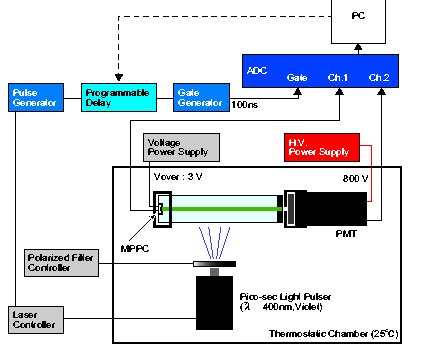}
\caption{\label{fig:tohru11}\em Sketch of the laboratory set-up used to determin the MPPC response function to light.}
\end{figure}

The MPPC response has a saturation behavior according to its intrinsic
property. Thus a saturation correction for each MPPC must be implemented.
We have measured the saturation effect of the MPPC using a simple bench test
in advance to perform the beam test. Figure~\ref{fig:tohru11} shows the setup 
of the response curve measurement using pico-second laser system, 
the scintillator-strip and the MPPC.


The measured MPPC response function is modeled with the function,
$N_{fired} = N_{pix} ( 1 - exp ( - eN_{in} / N_{pix} ))$, 
where $N_{fired}$ is the number of photons detected by the MPPC, $N_{pix}$ is the total number
of pixels on the MPPC, e is the photon detection efficiency and $N_{in}$ is the
number of photons incident in the MPPC sensitive area. From the fit to the curve in Figure~\ref{fig:tohru11}, one finds $N_{pix} = 2424 \pm 3$. The actual number of pixels in the MPPC device is 1600. The higher value found for $N_{pix}$ is due to the fact that photon generation in the scintillator has a time duration which allows some pixels to recover and fire a second time.
After applying saturation correction according to our own measurements the linearity of the electron energy response in the range 1-32 GeV is better than 1\%.

\subsection{Further plans for R\&D and integration}
The results obtained during the three beam test campaigns and with 
test bench studies indicate that the Sci-W ECAL fulfills the requirements 
for a particle flow detector. 
However, further improvements are expected
by introducing MPPC with more pixels, and scintillators with  smaller strip
width.
Our further R\&D for the coming years (2010-2012) will cover the following points:\\
(1) software effort: integrate Sci-W ECAL in the reconstruction and analysis framework of CALICE. \\
(2) integrate of electronics inside the sandwich layer structure,\\
(3) increase the scintillator segmentation to 5 mm wide strips,\\
(4) increase the number of pixel in the MPPC.

\section{Digital ECAL: DECAL}
\subsection{Concept}

Studies based on the concept of a digital ECAL (DECAL) started in 2006.
The groups involved so far have all been based in the UK and, due
to the serious funding difficulties for ILC work there, the project has 
been
limited in effort. Since 2008, the groups involved in the DECAL work 
are also
members of the UK SPiDeR collaboration, as the sensor implementation is 
likely to use Monolithic Active Pixel Sensors (MAPS) in CMOS. 
Hence, the technological aspects
of the sensor design are covered in the SPiDeR submission to this
PRC meeting~\cite{SPIDERPRC}.

The basic idea of a DECAL is to use very fine granularity
silicon sensors with binary pixel readout
for the active layers of a silicon-tungsten ECAL,
so as to make
an estimate of the number of charged particles in each layer.
These sensors would be used in place of the analogue ECAL silicon wafers
but otherwise the aim is that other aspects of the ECAL design,
such as the mechanics, would be unchanged. This would allow the DECAL
to be a ``swap-in'' option for a silicon-tungsten ECAL. Note, it also
means such a sensor would need to have a similar power consumption to
the analogue ECAL.

In order to count charged particles with binary readout, then the
pixel size must be much smaller than the separation of
particles within the showers so that the probability of two particles
hitting the same pixel is kept small. According to simulations,
the density of charged
particles in the innermost core of the highest energy electromagnetic 
showers expected for a LC may have a tail reaching to around 
100\,mm$^{-2}$.
However, high energy electromagnetic showers have
not been measured accurately at a high level of granularity.
Knowledge of the actual shower density
is clearly critical for the optimisation of the pixel size and this
is one of the major initial aims of the DECAL work.
For the studies performed so far,
a pixel size of $50\times 50\,\mu$m$^2$ has been assumed. This would 
require
an ECAL for a LC detector to have  $\sim10^{12}$ pixels.

\subsection{Motivation}
\begin{figure}
\centering
    \includegraphics[width=0.6\textwidth]{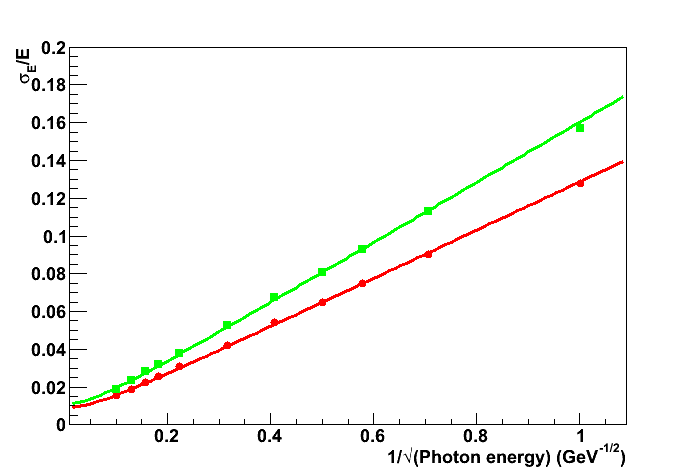}
\caption{\label{fig:DecalRes}
\em Simulated resolution of electromagnetic
showers in ideal analogue (green) and digital (red) ECALs with the same 
geometry.
The lines show the results of fits to the points of the form 
$\sigma_E/E = a \oplus b/\sqrt{E\ {\sl in\ GeV}}$ which give $a=0.9$, 
$b=0.128$
for the digital case and $a=1.1$, $b=0.160$ for the analogue case.}
\end{figure}

One basic motivation for a DECAL is that there is potentially a 
significant
improvement in the electromagnetic shower resolution. 
ECALs usually work on the principle that the number of
charged particles passing through each layer is on average proportional to
the incident particle energy; there are fluctuations around this
average due to the stochastic nature of the shower development.
These charged particles lose energy in traversing the sensitive
layers of the ECAL and the energy loss per particle also has
fluctuations, mainly due to the Landau distribution of energy
deposited but also due to variations in their speed and path length
through the material. An analogue ECAL therefore has contributions
to the energy resolution both from the intrinsic shower fluctuations
and also from the variations of the energy per charged particle.
The basic idea of a DECAL would be to reduce (or eliminate) the
latter contribution by attempting to measure the number of charged
particles directly. Figure~\ref{fig:DecalRes} shows an example of
the fundamental
resolution of electromagnetic showers simulated with an analogue
and a digital calorimeter of the same geometry. In this example,
the ECAL had 20 layers of tungsten of thickness $0.6X_0$ 
followed by 10 layers of thickness $1.2X_0$ for a total of $24X_0$;
these are typical of LC designs. In both cases, the resolutions
shown are ``ideal'', meaning they are based on perfect information
with no dead regions, electronics noise, threshold suppression etc.
It is thought that a realistic analogue ECAL will approach this idealised
value and the purpose of the DECAL studies is to determine the degree
to which this is true for the digital case. Simulation studies so
far indicate a degradation in the stochastic term from 0.128 to
around 0.14; this is still better than the ideal analogue case. A major
uncertainty in this more realistic resolution is the core shower density, 
as mentioned above; hence the need to measure this directly.

\begin{figure}
\centering
    \includegraphics[width=0.4\textwidth]{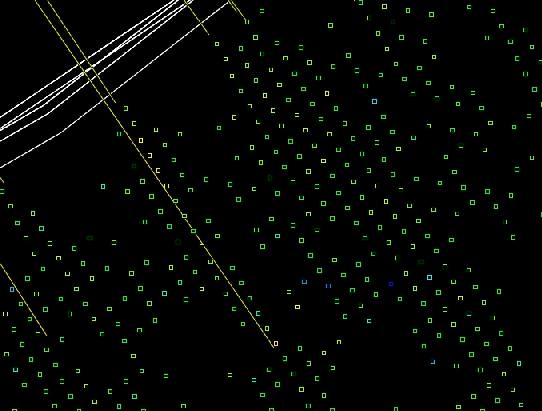}
    \includegraphics[width=0.4\textwidth]{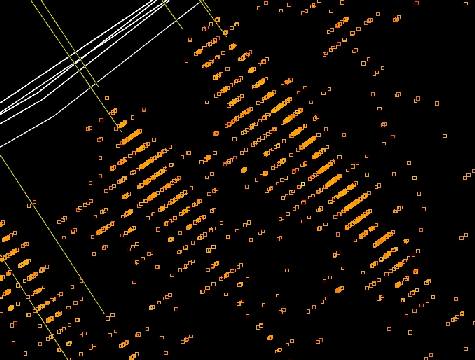}
\caption{\label{fig:DecalPFA}
\em Simulation of the same hadronic jet in the SiD detector with
(left) an analogue ECAL with 16\,mm$^2$ hexagonal pads and (right)
a digital ECAL with $50\times 50$\,$\mu$m$^2$ pixels.}
\end{figure}

The LC detector calorimeter designs are mainly driven by PFA performance,
rather than electromagnetic resolution; the latter plays only a small role 
in
the jet energy resolution which is more
significantly affected by the ``confusion
term'' due to mis-association of charged tracks and calorimeter deposits.
Systematic studies of the effect of having a very highly granular DECAL
rather than an analogue ECAL on the PFA performance have not yet been
done. However, figure~\ref{fig:DecalPFA} shows visually the difference of
the two in a simulated ILC hadronic jet.
It seems clear the much improved position resolution (from
around 5\,mm to 50\,$\mu$m) should not degrade the PFA
performance and can only help in untangling the detailed jet structure.

A further motivation for these studies is the ECAL cost. A 
silicon-tungsten
ECAL for a
LC detector is likely to require around 2000\,m$^2$ of silicon wafers
in total. The cost of these silicon wafers is of course an extremely
important factor. The analogue ECAL requires high resistivity wafers
which are not otherwise very widely used. This limits the number of 
vendors who will fabricate such wafers. In contrast, the aim is to 
make DECAL sensors using very standard CMOS processes which are widely
available. Having a large number of vendors for the sensor production
should help keep the price as low as possible.

\subsection{The TPAC beam test at CERN}
\begin{figure}
\centering
    \includegraphics[height=4.0cm]{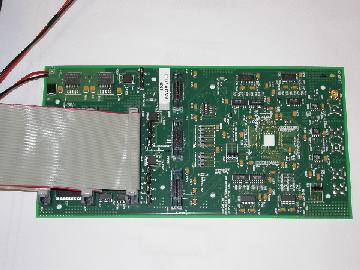}
    \includegraphics[height=4.0cm]{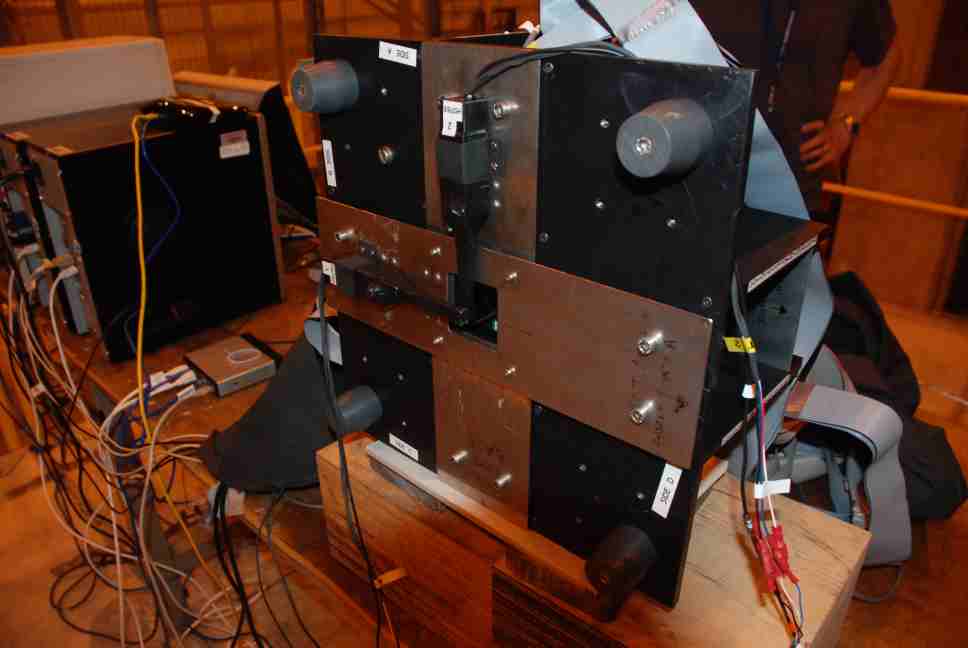}
\caption{\label{fig:DecalStack}
\em Left: One layer of the stack, showing the sensor mounted on the
PCB as the small silver square.
Right: The mechanical structure holding all six layers installed
on the beam line. Two scintillators to act as triggers are visible
in front of the beam entry hole.}
\end{figure}

The TPAC sensor was fabricated to study the issues of a DECAL sensor
for a LC detector. A stack of six layers with one
TPAC sensor per layer was operated at CERN in August this year mainly
for detailed performance measurements with MIPs.
Figure~\ref{fig:DecalStack} shows the equipment at the beam area.
Details of the TPAC sensor and the results from these MIP measurements 
are discussed in the SPiDeR submission~\cite{SPIDERPRC}. 

\begin{figure}
\centering
    \includegraphics[height=4.0cm]{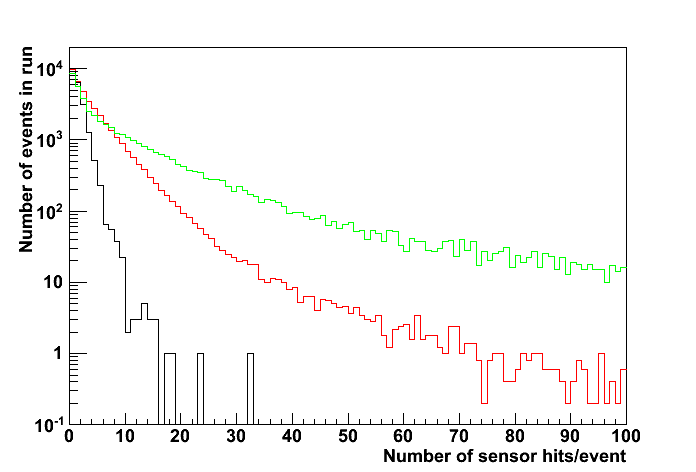}
    \includegraphics[height=4.0cm]{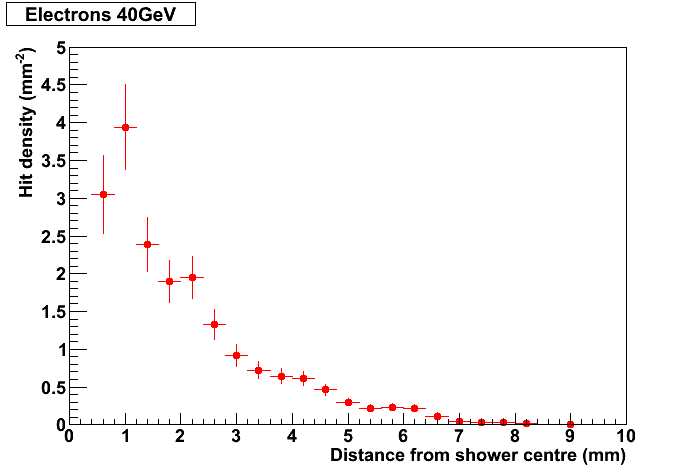}
\caption{\label{fig:DecalShower}
\em Left: Number of hits per sensor in time with the trigger 
measured at the CERN beam test in 2009 
for pions without tungsten (black), pions with tungsten (red)
and electrons with tungsten (green).
Right: Density of hits near the centre of electromagnetic showers
as a function of the
distance from the reconstructed shower centre. The data
are for an electron beam energy of 40\,GeV.}
\end{figure}

This sensor was not
intended for full shower reconstruction as it has an active area
of only $8.4\times 8.4$\,mm$^2$, smaller than the Moli\`ere radius
of tungsten. Hence running with electron beams was not an intended
goal of this beam test, particularly as the TPAC stack was running
parasitically to other main users who did not require electrons.
However, an unexpected opportunity for one day arose
and the stack was exposed to electron beams. 
For these runs, a 30\,mm stack of tungsten was added in front of
the first sensor layer. This corresponds to $8.6X_0$ of material,
meaning the particles emerge at close to shower maximum. There was
no further tungsten between the six sensor layers. These data
allow a first look at the core densities of electromagnetic showers.
Figure~\ref{fig:DecalShower} shows the much larger number of hits
seen in time with the trigger for the electron data than with
pions.

The shower centre was reconstructed by taking the average position of
all pixel hits in time with the trigger in the last five layers.
By comparing the
centre found using various combinations of layers, the resolution on
the shower centre using the five layers was found to be around 0.8\,mm.
The shower centre
was extrapolated to the first sensor layer and the density of hits
in this layer relative to the reconstructed centre was found.
Figure~\ref{fig:DecalShower} also
shows the average density distribution for 
electron showers. Note, the dense core of
the shower is averaged out to some extent
by the resolution of the shower centre and the tail is truncated by
the size of the sensor.

\subsection{Future plans}
In the short term, it is critical to make further measurements of
the core shower density. It is clear external tracking would help
locate the shower centre much more accurately and so a further
beam test early in 2010 is planned at the electron beams at
DESY where the EUDET telescope will be available. The DESY electron beams
are limited to 6\,GeV so another beam test at CERN may be undertaken,
if a slot is granted, later in 2010.

Longer term, a larger sensor will be produced to do a ``physics
prototype'' measurement of electromagnetic showers in a full-depth
DECAL stack. The development of this sensor is part of the SPiDeR
program and the technical aspects are discussed in that
submission~\cite{SPIDERPRC}.
The schedule is to produce a 30 layer DECAL stack in 2012 and expose
it to beam in a lengthy campaign that year, of equivalent scope to that
of the analogue ECAL in 2006/7.

The main aim of this program is a proof-of-principle of a DECAL.
These measurements should also give valuable physics information on 
electromagnetic showers, useful for tuning simulations.
In addition,
the data will give critical information required for the optimisation
of a real LC detector digital ECAL sensor in the future.
Specifically, the pixel size is a critical factor;
if the DECAL sensor measurements show the pixel size can be relaxed to
e.g. $100 \times 100$\,$\mu$m$^2$, then the power consumption would be 
directly reduced by a factor of four. The effect of noise on the 
resolution
should also be understood so slowing the pixel response (to again save 
power)
may then be an option if integrating over a few bunch crossings can be
shown not to degrade the performance. Another issue is that
the on-sensor memory storage causes
dead area so the importance of reducing this, potentially using smaller
feature size processes, compared with the risk of filling memory and
hence losing hits, will be clearer.

\section{Analogue HCAL: AHCAL}
\label{sec:AHCAL}

\subsection{Performance of the physics prototype and operational 
experience}
The analog hadron calorimeter prototype (AHCAL)
consists of a 38-layer sandwich structure of steel plates (4.5~$\lambda$
and  fine-segmented
scintillator tiles ($3\times 3$~cm$^2$ in the core region)
that are read out by wavelength-shifting fibers coupled
to SiPMs. There are 7608 channels in total.

\begin{figure}
\centering \includegraphics[width=0.48\textwidth]{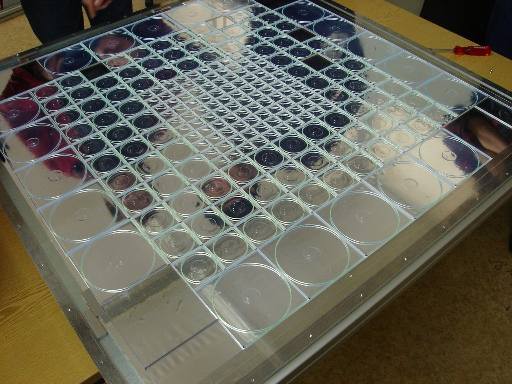}
\includegraphics[width=0.48\textwidth]{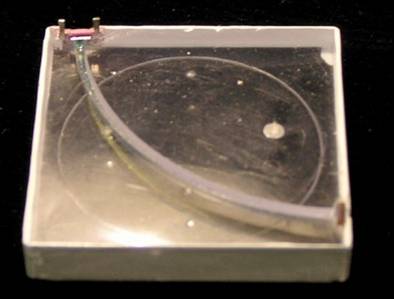}
\caption{\label{fig:FNALTB}\em (Left) AHCAL active module. (Right) Scintillator tile with wafelength shifter fiber and SiPM readout. 
}
\end{figure}

The signal is amplified and shaped using a modified version of the ASIC of
the Si-W ECAL prototype. A calibration/monitoring
system based on LED light monitors the SIPM gain and 
can measure the full
SiPM response curves in order to cross-check the non-linearity corrections 
derived from test bench data. 

The prototype has been operated 
for four consecutive years 
at two different test beam sites.  
The bias voltage of the detector 
is adjusted to set nominal light yield 
to 15 pixels/MIP for all channels. 
The average noise level per cell during the different test beam campaigns has 
been around (1.3-2) 10$^{-3}$  as expected from the test bench, but 
about a factor 10 higher than the final design goal. 
This is the first time that such a large quantity of SiPMs
is operated over longer periods. 
The detector 
had an initial number of dead channels corresponding to 2\% of the total. 
An additional 0.5\% of channels died during transportation to FNAL. 
The large majority 
of these  are due to bad soldering or subsequent broken connection.
A detailed study of SiPM performance over two years of operation ('07-'08) 
reveals that 
this number has been very stable during operation. 
Only 5 SiPMs died during operation, possibly due to too high current 
induced by surface defects.
The SiPM used 
in the CALICE AHCAL thus demonstrated excellent stability 
during four months of operation.  

\subsection{Calibration and stability}
The calibration scheme for the conceived scintillator HCAL is based on
test bench characterizations and test beam data. While electromagnetic,
hadronic and weighted energy scales can be established with sample
structures of the HCAL alone or in conjunction with the ECAL exposed to
beams of muons, electrons and hadrons, the equalization of the
detector cells must be established with muon beams for all active layers
of the detector individually.
Due to the proposed deep underground site, the power pulsed electronics
and the fine segmentation, cell equalization cannot be repeated using cosmic rays.

The calibration accuracy is maintained using LED monitoring of the
photo-sensor gain, in-situ MIP calibration based on track segments in
hadron showers and classical slow-control recording of the relevant
operation parameters, temperature and bias voltage. These methods have
been successfully applied to test beam data.

Simulating ILC events and using algorithms bench-marked with test beam 
data,
we have quantitatively determined the required luminosity for in-situ
MIP calibration of individual cells and of average values for sub-sections
of the detector. A cell-by-cell
MIP calibration is not possible with realistic running times, but it is
also not necessary.
Using fully detailed simulations of the ILD detector and reconstruction
based on the Pandora particle flow algorithm~\cite{PFA}, we have modeled different
scenarios of statistically independent as well as coherent mis-calibration
effects, affecting the entire HCAL or parts of it.
Purely statistical variations, like those arising from calibration errors
or random aging effects, hardly affect the hadron energy resolution at 
all,
due to the large number of individual cells contributing (10 per GeV).
Coherent effects which could for example arise from uncorrected 
temperature
variation induced changes of the response are potentially more harmful
if they affect a large portion of the detector.
However, these are also easy to detect,  and propagate only mildly
into the jet energy resolution.

\begin{figure}
\centering
    \includegraphics[angle=0,width=0.5\textwidth]{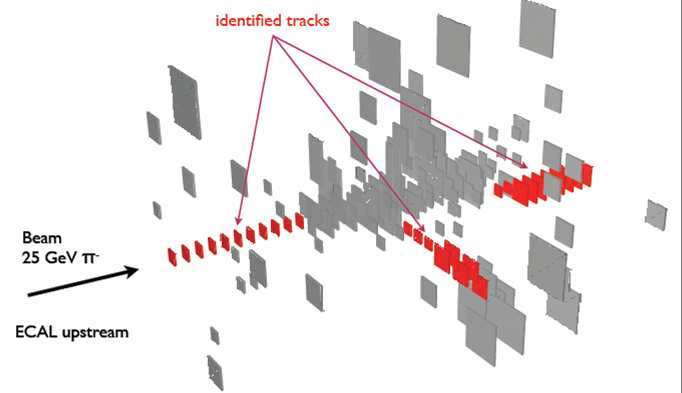}
    \includegraphics[angle=0,width=0.45\textwidth]{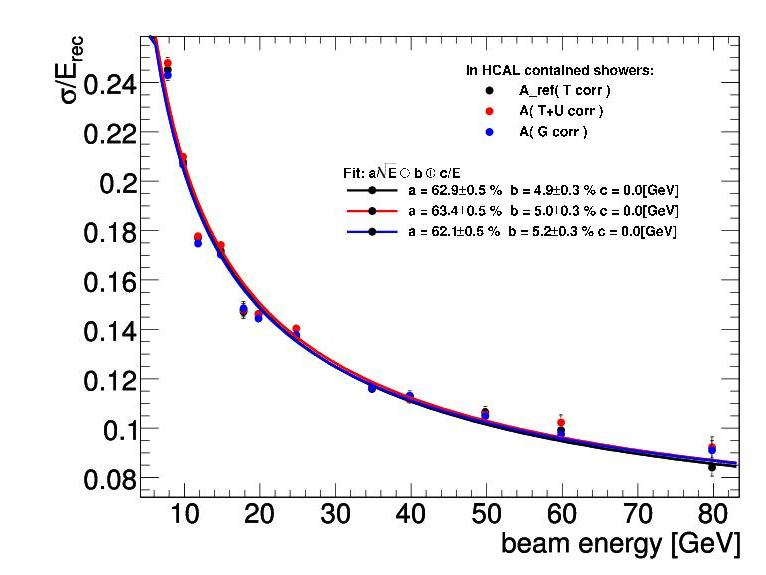}
\caption{\label{fig:calibreso}\em Event display of a hadron showers with
identified track segments uses for in-situ MIP calibration.
Hadron energy resolution obtained with MIP calibration from the same
and from a different test beam site, transported with in-situ methods 
only.}
\end{figure}

We have demonstrated the validity of these simulation based~\cite{CAN-018} estimates
by treating our test beam experiment like a collider detector, using
cell-by-cell inter-calibrations only from data taking at a different site,
under different conditions and after having it exposed to disassembly,
transport and re-assembly influences. Applying only in-situ monitoring techniques, we re-established the 
scale and reproduced the resolution;
see the right plot in Fig.~\ref{fig:calibreso}. The constant term is larger here than
reported in the analysis section due to a preliminary set of calibration
constants used.
Imperfections, due to intermittent hardware malfunction, and absent in any
simulation showed up, but were successfully compensated.
This is an important outcome of the test beam runs at CERN and at FNAL;
it gives confidence that the high granularity approach with the novel SiPM
technology can be extrapolated to the full detector scale.

\subsection{Development of a technological prototype}
In order to validate the concept of a highly granular scintillator based 
HCAL,
it needs to be demonstrated that the high channel density can actually be
realized without compromising the performance by too many dead spaces or
reduced compactness and hermeticity once readout and calibration
electronics or support structures have been accommodated.
In this respect the physics prototype is not
scalable and needs to be complemented by a technological prototypes
addressing these integration issues.

\begin{figure}
\centering
    \includegraphics[angle=0,width=0.7\textwidth]{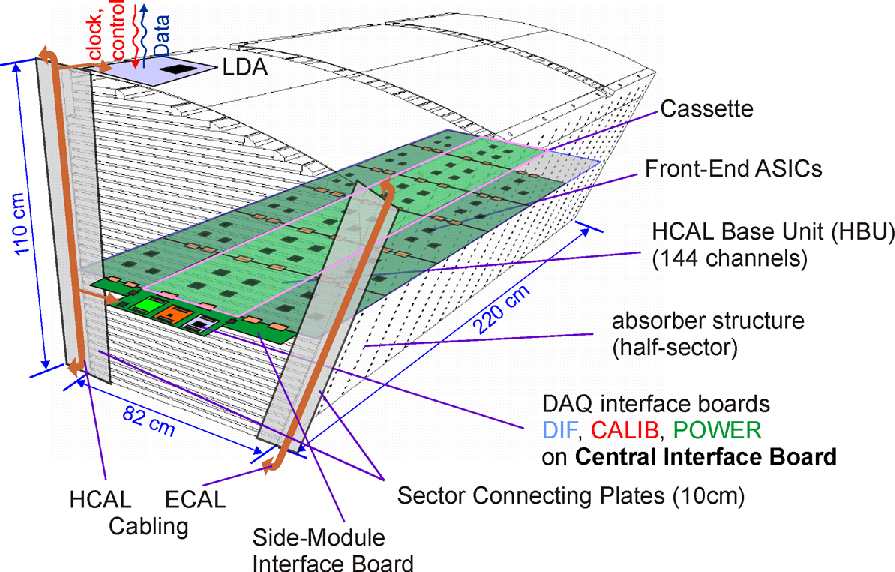}
\caption{\label{fig:ahcalmodule}\em Electronics integration
architecture for the technological AHCAL prototype.}
\end{figure}

The envisaged detector architecture~\cite{EUDET_memo} is sketched in 
Fig.~\ref{fig:ahcalmodule}.
It is inspired by one variant of the ILD detector concept, but is very 
similar
to those envisaged for SiD or CLIC. The Figure shows one sector of a
barrel subdivided only once along the beam axis. This layout provides
access to electronics and service interfaces once the detector end-cap is
opened, but poses tight space constraints to the barrel end-cap
transition region.

Compared with existing hadron calorimeters, the particle flow HCAL has a 
rather
fine longitudinal sampling, with a correspondingly high pressure on the 
thickness
of the active layer gaps, but also on mechanical tolerances.  This, 
together with the
requirement of minimum dead zones represents an engineering challenge
which is being addressed now.

There are 48 independent read-out layers which must be as thin as
possible in order to keep the overall detector volume small,
as it has to fit inside the main solenoid of the collider detector and
still provide maximum hadronic absorption depth. Each layer has a fine
transverse segmentation, again with individual cell read-out,
which requires concentrating the data at an early stage in order
to keep connectivity issues manageable and reduce dead areas
occupied by external electronics components.

The readout ASICs process the signals including digitization and
intermediate storage. Thanks to their operation with pulsed power
synchronous with the accelerator time structure, no cooling inside
the modules is necessary. Only interfaces to DAQ, power supplies
and calibration control are placed outside the volume.

\begin{figure}
\centering
    \includegraphics[angle=0,width=0.45\textwidth]{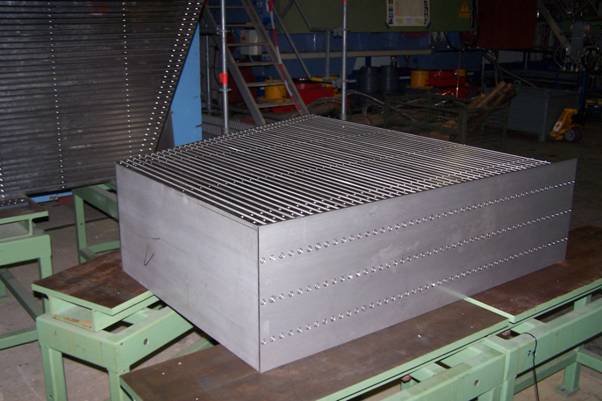}
    \includegraphics[angle=0,width=0.45\textwidth]{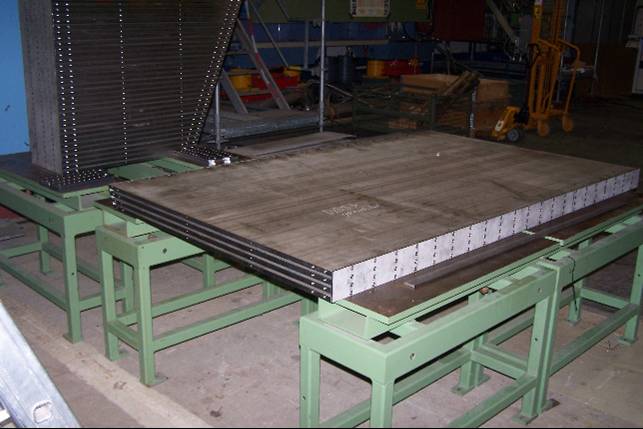}
\caption{\label{fig:hcalstructure}\em The two vertical (left) and 
the horizontal (right) mechanical structures for the technical HCAL prototype.}
\end{figure}

We foresee a staged prototype program to validate this concept:
\begin{enumerate}
\item A horizontal test structure (Fig.~\ref{fig:hcalstructure} right) to establish the mechanical tolerances
and electronics signal integrity over the full area and length of the 
module.
Heat dissipation and installation procedures can also be studied in full 
scale.
\item A vertical test structure (Fig.~\ref{fig:hcalstructure} left) to establish the mechanical stability
under various orientation and stresses as well as multi-layer electronics
integration and operation.
This can be tested with electron beams and requires instrumentation of
a small volume only.
\item  The vertical structures can be stacked in a flexible way, according
to the needed overall test beam geometry, and their instrumentation
completed for hadron shower studies.
\end{enumerate}
The first two steps can be accomplished with about one to two thousand 
channels,
the third will require about  40'000 channels. The multi-layer set-ups
need a compact realization of the interface electronics components.

In addition to the demonstration, that the mechanical and electronics 
design
challenges can be met, there are operational and physics issues to be 
studied,
which could not be tackled with the physics prototype:
\begin{itemize}
\item Establish stable operation with auto-triggering and on-detector
zero-suppression. This will require continuous monitoring of thresholds
and in-time adjustment of bias voltages to compensate for temperature
induced variations of the signal.
\item The new ASICs incorporate a TDC for a time measurement of 
calorimeter
hits. It must be shown that this system can be timed in and operated 
stably.
\item The usefulness of the timing measurement for neutron hit
identification  and shower reconstruction can be evaluated.
\item The time evolution of hadron showers in simulation models
can be confronted with experimental data; for the first time in a 
spatially
resolved way.
\item The stainless steel structure allows for operation of the 
calorimeter
in a magnetic field,  and to test the predicted influence of the field on 
hadron
shower propagation.
\end{itemize}

\subsection{Mechanical structures}
The module consists of stainless steel plates with 16~mm thickness which
are supported only by the 5mm thick side panels, without additional 
spacers.
The gaps can be filled completely with active elements.

The mechanical stability of this ambitious design has been extensively
studied using finite element methods. Deformations of the individual 
module
as well as the overall barrel structure were validated, taking external 
support
and the suspension of the ECAL into account.

It is straightforward to keep the required tolerances by precisely
machining all absorber plate surfaces. However, the cost would be higher
by a factor 2 to 3 with respect to the raw material. Since cost 
optimization
is an important ingredient in a "realistic" structure, we aim at producing 
the
structures from raw, commercially available rolled plates. In particular 
the flatness
of the plates, as specified by commercial standards, exceeds the 
requirements by
intolerable amounts. We have prepared devices for measuring deviations and
found a cost-effective solution (roller leveling) to achieve the required
flatness within a millimeter across the full module.

The horizontal and two vertical test structures are shown in 
Fig.~\ref{fig:ahcalstacks}.
The first has 4 plates, 2160~mm long, the second is 360 mm deep.

\subsection{Readout and calibration electronics}
The electronics is subdivided into HCAL Base Units (HBUs) in order to keep
the single modules at reasonable sizes concerning production and handling.
The HBU with a typical size of $36\times36$~cm$^2$ integrates 
144~scintillating
tiles each with Multi-Pixel Geiger-Mode Photo-diodes (MGPDs). The analogue
signals of the MGPDs are read out by four front-end ASICs (SPIROC, [3]). 
The full
depth of a sector is covered by 6 HBUs that are connected together by so 
called
interconnection flexleads. A chain of 6 HBUs forms an electrical layer 
unit.
 the so called slab. At the end of the sector, the electronics of a 
detector layer
are connected to the detector interface module DIF and the AHCAL specific
modules for calibration (CALIB) and the power-supply module (POWER).

\begin{figure}
\centering
   \includegraphics[angle=0,width=0.45\textwidth]{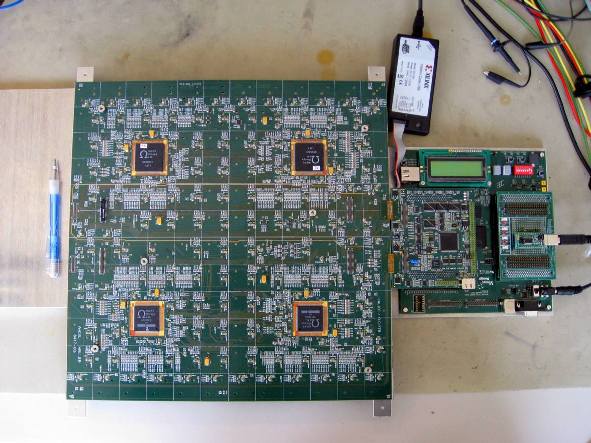}
    \includegraphics[angle=0,width=0.45\textwidth]{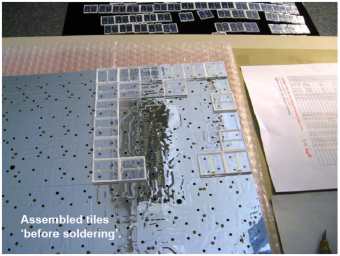}
\caption{\label{fig:hbu-test}\em Test bench set-up with the prototypes of 
the HBU and
interfaces (left), back-side of the HBU with reflector foil and tiles.}
\end{figure}

In order to keep the development time for the first prototype modules as 
short as
possible, the first DIF module is realized by a commercial FPGA board.
The modules CALIB and POWER are realized as mezzanine modules on top of 
the
DIF, while the interface to the first HBU prototype is realized by the 
final thin
flexlead interconnections via the CALIB module.
The prototype setup shown in Fig.~\ref{fig:hbu-test} can be used
to test all electrical characteristics of a full layer in the horizontal
test structure.
The electronic commissioning has been completed using LED signals, and the
first HBU prototype will be exposed to the DESY test beam in November.
The DIF board will be read out via USB; the final DAQ chain will
be commissioned in 2010, this can start with the FPGA board.
There are still two SPIROC chips of the first version of the HBU, which 
have a
thicker package, therefore a re-design of the board layout will be needed
for the full layer test.

The AHCAL specific module CALIB is used to control and operate the light
calibration system that is based on electrical signal distribution (see 
below) as
well as a charge -injection system that can be used for calibration and
debugging as well. The POWER module provides the supply voltages for the
 inner-detector electronics as well as for the modules DIF and CALIB, and 
the
MGPD bias voltage.  The architecture of the power regulator setup has been
defined in detailed measurements of the power-pulsing characteristics.
A settling time below 10~$\mu$s can be expected.

For the LED calibration and monitoring system we follow two approaches,
one based on a central driver and optical signal distribution, one based 
on
electrical signal distribution and an individual LED per tile. The latter
is already integrated in the HBU, but not yet fully optimized. The aim
is to equalize the light intensity and maximize its dynamic range.
The former distributes light by a notched fibre to a row of tiles.

\subsection{Scintillator and photo-sensor development}
Following the successful operation of the physics prototype,
progress was made by various manufacturers, e.g.  in Russia or Japan,
to provide sensors with lower dark count rate and / or smaller inter-pixel
cross-talk which allow to decrease the noise occupancy above threshold
of $10^{-3}$ in the present prototype by an order of magnitude and thus
fulfill the requirements from both physics  (for neutron hit 
identification) and
DAQ band width. They will be used in the technological prototype.

For the coupling of sensors to scintillator and PCB different approaches
are being followed, based on either wavelength-shifting WLS fiber mediated
or direct read-out, which becomes possible with blue-sensitive 
photo-diodes.
In the direct coupling case, the sensor is mounted either in SMD style 
with its
sensitive surface in the PCB plane, or in the same position as for
tiles with fibres. In both cases it collects the scintillation light
directly from the tile, which has to be shaped in a dedicated way to
compensate for the otherwise prohibitive light collection 
non-uniformities.
In this case the positioning of the sensor with respoct to the scintillator is somewhat less 
critical.

The different schemes can be accommodates and tested with only modest
modifications of the HBU design. Similarly, we plan a version for the
scintillator strip ECAL which has twice the channel density.

\subsection{Tungsten absorber}
Simulation and reconstruction studies using the Pandora algorithm
have indicated that particle flow is a promising approach also for
a linear collider in the multi-TeV range. As the jet energies increase,
not only the particle separation becomes more difficult (confusion),
but also leakage is an increasingly important contribution to the
energy resolution. A hadron calorimeter for high energies must have
a depth of at least 7~$\lambda$. Since it would be difficult and costly to
increase the solenoid radius,  tungsten as absorber material promises
to offer a competitive solution.

The simulations suggest that a sampling structure with 10~mm thick 
absorber
plates and 5~mm scintillator would optimize the resolution within
the given spatial constraints.

In order to validate the simulations, which have other sources of
uncertainty than those for steel, a tungsten absorber structure should
be studied with test beams and finely segmented readout (scintillator and 
gas).
We plan to exchange the steel plates of the physics prototype by tungsten
plates and perform first studies with the existing scintillator layers, 
followed
by gaseous readout as it becomes available.

Due to the enhanced role of neutrons in the tungsten calorimeter response,
as compared to steel, and because of the need for an extremely compact
design, we aim at studying tungsten also with the readout structures
of the technological prototype, which provide time measurements.

\section{Digital HCAL: DHCAL}
\subsection{Description of the project}

The digital hadron calorimeter project trades the typical tower structure
of past hadron calorimeters and their high-resolution readout with large
number of pads read out individually with a single bit resolution. A
collaboration of U.S. institutes (Argonne, Boston, FNAL and Iowa) is
developing such a novel calorimeter with Resistive Plate Chambers (RPCs)
as active media. Currently the readout is segmented into $1\times 1$\,cm$^2$ pads 
or
10,000 per square meter.

The project progresses in several stages. In a first stage different
designs of RPCs were developed and tested with a high-resolution readout
system. In parallel to this activity a 1-bit readout system capable of
handling large numbers of channels was developed. The second stage put the
two together in a small prototype calorimeter, here named the Vertical
Slice Test (VST), and included detailed tests with both cosmic rays and in
the FNAL test beam. For the first time within the CALICE collaboration
the VST utilized a readout system with the digitization taking place
directly on the front-end boards. Based on the successful experience with
the VST and after a further round of R\&D, the third stage consists of the
construction of a larger prototype hadron calorimeter with close to
400,000 readout channels. The calorimeter will be tested extensively in
the FNAL test beam. The following stage will return to R\&D to tackle
the remaining technical issues before embarking on the construction of a
so-called technical prototype.

Additional information pertaining to this project can be obtained from
\cite{jose1}.

\subsection{Past achievements}

Detailed measurements \cite{jose2} with an analogue readout system were 
performed.
The results served as basis for the design of RPCs, optimized for the use
in calorimetry, and the corresponding electronic readout system. Of
particular interest is our development of a novel chamber design which
utilizes only a single glass plate.

\begin{figure}\centering    
  \includegraphics[width=0.4\textwidth]{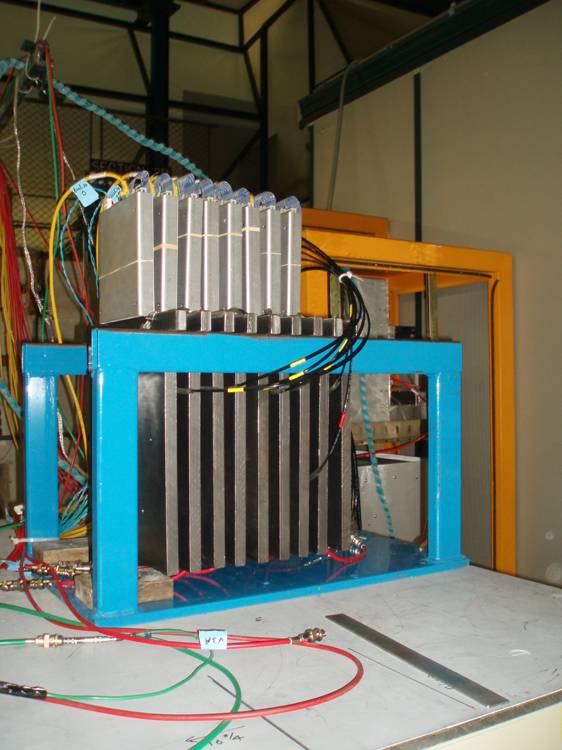}
  \caption[]{\label{jose_fig1}\em Photograph of the Vertical Slice Test in the 
FNAL test beam.}
\end{figure}

The VST was extensively tested with cosmic rays and in the FNAL test
beam, see Fig.~\ref{jose_fig1}. In a first round the calorimeter was 
exposed to a
broadband muon beam. The data were utilized to measure the MIP detection
efficiency and the average pad multiplicity.  The results were published
in \cite{jose3}.

\begin{figure}
\centering
\includegraphics[width=0.45\textwidth]{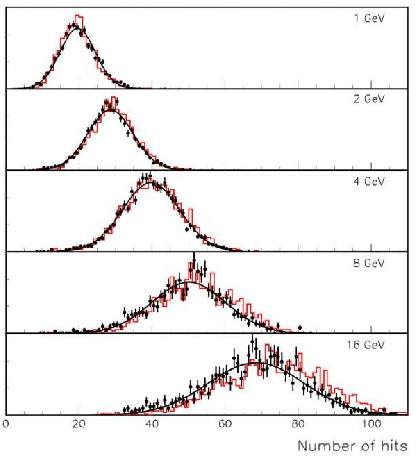}
\includegraphics[width=0.48\textwidth]{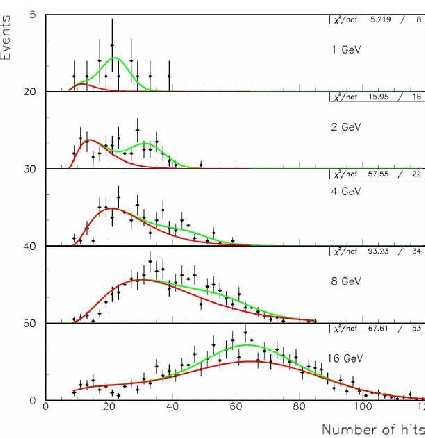}
\caption{\label{fig:DHCAL}\em  Response (i.e. number of hits) 
for a selection of  beam energies for electrons 
(left) and pions (right) incident on the DHCAL test module.  
The results of a GEANT4 simulation are indicated in red.  
In the pion case it is necessary to take account 
of the residual contribution from positrons (shown in green). }
\end{figure}

The response to positrons and pions was studied for beam momentum settings
at 1, 2, 4, 8 and 16\,GeV/c. The distribution of number of hits and the
longitudinal and lateral shower shapes were measured \cite{jose4,jose5}. 
The results
were compared to a Monte Carlo simulation of the set-up based on GEANT4.
As an example 
Fig.~\ref{fig:DHCAL} 
shows the number of hits obtained with 
positrons at
the various beam settings. Notice the deficiency of hits at the higher
momenta, due to the limited rate capability of RPCs.


The rate capability of RPCs was measured utilizing the 120\,GeV primary
proton beam at different intensities \cite{jose6}. For rates above 
100\,Hz/cm$^2$ a
clear decrease in efficiency is observed. The group developed an
analytical model which adequately reproduces the observed effects.


\subsection{Current activities}

Currently the collaboration is assembling a large prototype calorimeter,
the physics prototype. The calorimeter will consist of 38 active layers
interspersed with absorber plates. The latter will be provided by the
Analogue Hadron Calorimeter (AHCAL) stage. Each layer will measure 
$96 \times 96$\,cm$^2$ and contain three separate chambers. With a segmentation of 
$1 \times 1$\,cm$^2$
laterally and layer-by-layer longitudinally, the calorimeter will count
close to 400,000 readout channels.

The group is actively assembling chambers. Whenever possible the tasks are
achieved with the help of jigs or fixtures. The spraying of the glass
plates is performed within a spraying booth and is entirely automatic. The
frames of extruded PVC are cut and predrilled (to allow for the insertion
of the fishing lines, acting as spacers between the glass plates) using a
specially designed jig. The RPCs are assembled within a fixture providing
exact dimensions at the 25 - 50 $\mu$m level.

The front-end ASICs, the DCAL III chips, have been produced and over
10,000 units are in hand. A handful of chips underwent extensive testing.
No design faults have been discovered. The DCAL III chips are currently
being tested by a robot at FNAL. Apart from the first wafer which
showed a low yield, the average yield appears to be in excess of 90\%. So
far, the robot tested approximately half the 10,000 chips.

\begin{figure}\centering    
  \includegraphics[width=0.7\textwidth]{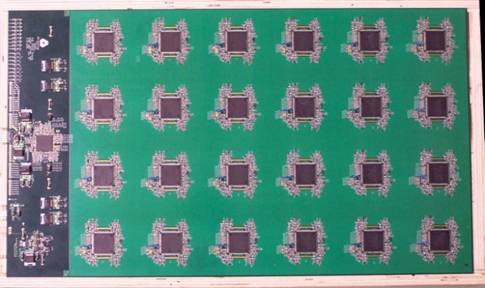}
  \caption[]{\label{jose_fig3}\em Photograph of the front-end board containing 
24 DCAL III chips.}
\end{figure}

The front-end boards contain each 24 DCAL chips and cover an area of 32 x
48\,cm$^2$, see Fig.~\ref{jose_fig3}. The boards also contain the data 
concentrator seen in
darker green on the left side. The board has been prototyped, the firmware
written and debugged. The board was utilized to collect Cosmic ray data in
conjunction with large chambers. A second round of prototyping is
currently underway, before initiating mass production.

The data collector boards have been fabricated and assembled. Of the 35
boards about six have been tested so far. The performance is as expected.
The timing and trigger module is currently undergoing a re-design.
Prototypes of the new modules are expected within the next two months.

The high voltage system is in hand and its computer controlled program has
been commissioned. A gas mixing rack has been assembled and successfully
tested. The gas distribution rack utilized in the VST will undergo a minor
extension to be used with the physics prototype. The low voltage power
supplies are in hand. Power distribution boxes have been designed and a
first prototype has been built.

To facilitate combined running with other CALICE calorimeter prototypes,
the DHCAL DAQ software has been written within the framework of the CALICE
DAQ software. The software is written and is currently used to collect
Cosmic Ray data with the new boards.

Finally, the performance of the physics prototype has been simulated in
great detail, including the actual response of RPCs \cite{jose5}.

The construction of the physics prototype is expected to be completed by
April 2010, to be followed by a first test period in the FNAL test
beam. Combined tests with the CALICE Silicon-Tungsten will take place in
the late summer of 2010. A final running period is foreseen for early
2011.

\subsection{Plans for the future}

Despite the large effort of assembling the physics prototype the
collaboration has initiated R\&D concerning the remaining technical issues
of an RPC-based hadron calorimeter. Table~\ref{jose_tab1} summarizes the 
various
(planned) activities.

\begin{table}[htbp]
  \centering
  \begin{tabular}[c]{p{4cm}|p{4cm}|p{8cm}}
R\&D topic	& Funds	                &Comment\\
\hline \\
Thin RPC	&Applied for	        &Further investigation of 1-glass 
design\\
Large area RPCs	&Currently not pursued	&Areas of several m$^2$ needed\\
Gas system	&Funded	                &Exploration of new gas mixtures, 
recycling, gas distribution\\
High Voltage distribution &Funded	&System capable of supplying HV to 
all layers of a module individually\\
Low Voltage distribution  &Currently not pursued	&System capable of 
supplying LV to all layers of a module individually\\
Wedge shape	&Currently not pursued	&Develop concept to accommodate 
wedge shaped module designs\\
Pad/FE-board	&Currently not pursued	&Develop new design which 
minimizes thickness\\
Front-end ASIC	&(Funded)	        &Develop next iteration with 
reduced power consumption, token ring passing, and redundancy for 
reliability\\
Data concentrator	&Currently not pursued	&Develop new system which 
minimizes space requirement and provides high reliability\\
Mechanical structure 	&Currently not pursued	&Develop cassette 
structure which can be oriented which ever way, develop module structure 
which accommodates all supplies and data lines\\
Magnetic field	&Currently not pursued	&Tests of all subsystems in 
magnetic field\\
  \end{tabular}
  \caption{\em Summary of R\&D topics beyond the construction of the physics 
prototype.}
  \label{jose_tab1}
\end{table}

\subsection{GEM DHCAL Status}       
\begin{figure}\centering    
\includegraphics[angle=-90,width=0.6\textwidth]{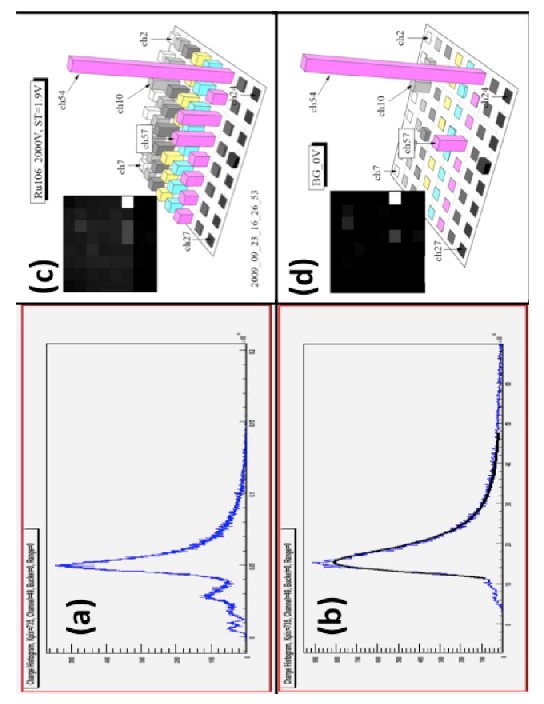}
\caption[]{\label{fig:gem1} a) Pulse height distributions from $\rm {Fe^{55}}$ 
 source, showing characteristic peaks from 5.9KeV and 4KeV X-rays  (b)Pulse 
height distribution from $\rm {Ru^{106}}$  source, conforming to a Landau 
distribution (c) Lego plot of hits in all active channels with radioactive 
source (d) Lego plot of noise hits without high voltage applied on the chamber}
\end{figure}

The University of Texas at Arlington HEP team has been developing a 
digital hadron calorimeter (DHCAL) using the Gas Electron Multiplier 
(GEM) as the sensitive gap technology.  The team has constructed several 
prototype chambers to date and has completed constructing a new chamber, 
GEM4, with the new gas-transparent G10 spacer from CERN and the updated 
KPiX readout board.  The team has been performing chamber characterization, 
using $\rm {Fe^{55}}$ and $\rm{Ru^{106}}$ radioactive sources.  Figure~\ref{fig:gem1}(a) 
shows two distinct peaks from $\rm {Fe^{55}}$ X-rays in one of the 64 
KPiX readout channel.  Figure~\ref{fig:gem1}(b) shows the Landau distributions 
from $\beta$-particles from a $\rm {Ru^{106}}$ source.  The team is now working on two 
dimensional measurement.  The sources were placed at a sufficient height to 
illuminate all active channels, demonstrating two dimensional profile distributions 
in Fig.~\ref{fig:gem1}(c).  The three channels (54, 57 and 10) with a large 
number of hits arise from electronic noise as demonstrated clearly in 
Fig.~\ref{fig:gem1}(d), which shows the scatter plot of hits without high voltage 
supplied onto the chamber - the noise distribution is the same even with HV.

The team is in the process of investigating the uniformity of the chamber responses. 
The team will determine chamber gains using sources and will take cosmic ray data 
for further MIP studies and efficiency studies.   

\subsection{Future Plans for GEM DHCAL}
In order to keep up progress toward a GEM based DHCAL, the UTA team plans a three-phased approach as described below.
\subsubsection{Phase I: Through Late 2009}
\begin{itemize}
\item {\it 30\,cm$\times$30\,cm Chamber Characterization}: Building on the 
successes accomplished in reading out multiple channels using 64 channel 
KPiX version 7 chip, the team will complete bench characterization of the 
chamber using radioactive sources and cosmic ray particles.   The team 
then will take the chamber to a particle beam and take data to complete 
characterization of the chamber.   At the same time, since the team plans 
to ultimately take data in the CALICE stack as a DHCAL, it will investigate 
the behaviour of the chamber using the DCAL digital 
readout chip jointly developed by ANL   and FNAL.
\item {\it 33\,cm$\times$100\,cm Large GEM Foil Development}: The team has 
been working with the CERN GDD workshop in developing 33\,cm$\times$100\,cm GEM 
foils, the largest to date, and finalized the design of 
33\,cm$\times$100\,cm GEM foils that have the active area dimension of 
32\,cm$\times$96\,cm and sent it to the CERN GDD workshop.   
The CERN GDD workshop 
has been successful in producing GEM foils with uniform hole shapes.  
The initial prototype run for 33\,cm$\times$100\,cm GEM foil was attempted 
in late August 2009 but the quality of the holes was unsatisfactory.   
CERN GDD has been working on refining the process and is ready for 
another run. 
\end{itemize}

\subsubsection{Phase II: Late 2009 -- Late 2010}
\begin{figure}\centering    
\includegraphics[angle=-90,width=0.6\textwidth]{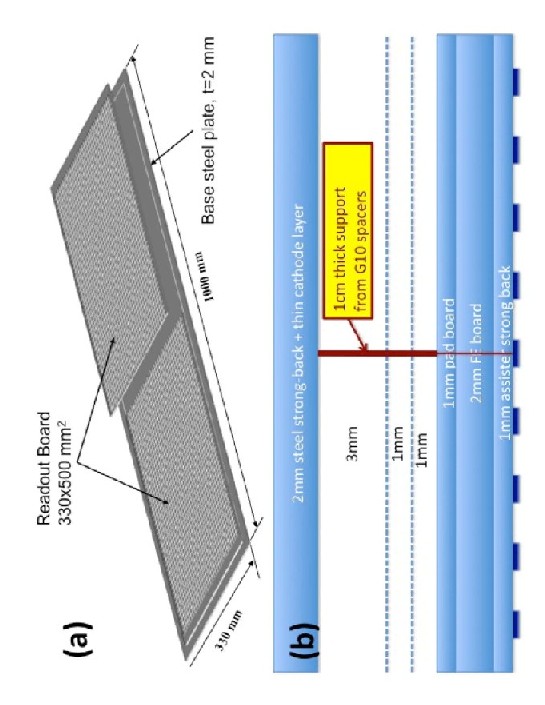}
\caption[]{\label{fig:gem2} (a) A 3D schematic diagram of a 33\,cm$\times$100\,cm 
unit chamber that has two 33\,cm$\times$50\,cm anode boards for readout. (b) Cross 
sectional view of a 33\,cm$\times$100\,cm unit chamber.  The spacer has a 1\,cm thickness
 to provide sufficient area for gluing two anodes boards in the middle.}
\end{figure}

\begin{itemize} 
\item {\it 33\,cm$\times$100\,cm Unit Chamber Development}: During this period, 
the team will certify GEM foils as they get delivered and will work on the assembly 
technique for 33\,cm$\times$100\,cm unit chambers.  
\item {\it Characterization of 1024 Channel KPiX Chips}:  The team plans 
to work on characterization of the 1024 channel KPiX chips with the intent 
to use them in 33\,cm$\times$\,100\,cm unit chamber characterization.  The latest 
30\,cm$\times$30\,cm chamber will be used for chip characterization before assembling 
them in large unit chambers.
\item {\it Understanding of Chamber Behaviour with DCAL Chips}:  The team will 
continue working on understanding the chamber behaviour with the DCAL chip, if this is 
not completed in the previous period.  
\item {\it Begin Construction of 33\,cm$\times$100\,cm Unit Chambers}: The team will
 begin construction of 33\,cm$\times$100\,cm unit chambers.   The team plans to build 
one using the most understood KPiX chip-based anode boards to fully characterize 
33\,cm$\times$100\,cm unit chambers.  The team will build fifteen additional chambers 
with DCAL chip based anode boards to use them in assembling the final 100\,cm$\times$100\,cm
DHCAL sensitive gap planes.  Figure~\ref{fig:gem2}(a) shows a three dimensional 
schematic diagram of the unit chamber.  It clearly shows the two readout boards 
of 33\,cm$\times$50\,cm since the flatness of the boards any longer than this size 
is not guaranteed. Figure~\ref{fig:gem2}(b) shows the cross section of a unit chamber.
   It shows that the total thickness of these chambers will be of the order of 11\,mm,
 including a 2\,mm steel plate that acts as a strong-back.  The spacer has a 1\,cm thick 
support in the middle of the chamber to provide sufficient surface area for the anode 
board to be glued on.   We also plan to put another 1\,mm thick strong-back support 
steel plates with holes that allows readout chips to protrude.  This steel plate 
does not add any additional thickness to the overall dimension of the chamber.
\end{itemize}

\subsubsection{Phase III: Mid 2010 -- Late 2011}
\begin{itemize}
\item {\it Unit Chamber Characterization with DCAL Chips}: The team will 
characterize 33\,cm$\times$100\,cm unit chamber built with DCAL chip based anode 
board on the bench and in particle beams.\item {\it Construction of Five 
100\,cm$\times$100\,cm Chamber Planes}: The team will complete construction of all 
fifteen 33\,cm$\times$100\,cm unit chambers and complete assembling five 
100\,cm$\times$100\,cm sensitive detector planes.  The team plans to put three unit chambers together 
on a steel plate to provide mechanical strength to the 100\,cm$\times$100\,cm 
detector planes. These planes will be inserted into the existing CALICE calorimeter 
stack and run together with Si/W ECAL and 35 RPC planes in the HCAL. One critical 
constraint in this beam test is that the CALICE 
stack can stay at in the USA only until April 2011 due to customs rules.  
\item {\it TGEMs and RETGEMs}: The team has also been looking into Thick GEM (TGEMs)
 and Resistive TGEMs (RETGEMs).  Weizemann Institute in Israel has been 
collaborating with us in developing large size TGEMs.   Thus, if these two new 
types of GEMs become available and if our resources allow, the team plans to 
explore the possibility of using these GEMs to produce chambers and beam test them. 
\end{itemize}

\section{Semi-digital HCAL: SDHCAL}
\label{sec:sDHCAL}
\subsection{Introduction}


Members of the CALICE collaboration,
including Belgian, Chinese, French, Russian and Spanish groups, are pursuing 
a new development aimed at constructing a highly granular gaseous hadronic 
calorimeter prototype based on a semi-digital readout and a transverse 
segmentation of 1\,cm$^2$. 
In addition to the tracking capability it offers, a semi-digital readout 
HCAL can provide very good energy resolution which can be, according to 
simulations, as good as that of an analogue calorimeter 
with an appropriate choice of 
threshold values. The semi-digital HCAL prototype is intended to come as 
close as possible to the hadronic calorimeters of the future ILC 
experiments in terms of resolution, efficiency and compactness.

Two kinds of gaseous detectors are being investigated as candidates 
to become the 
sensitive medium of such a SDHCAL: glass RPCs (GRPC) and Micromegas. New readout 
electronics satisfying the ILC constraints was developed and successfully 
tested on small GRPC detectors and more recently on a 1 m$^2$ detector.  
Tests of small and medium size Micromegas detectors were also performed,
yielding promising results. A 1 m~$^{2}$ Micromegas prototype is under
construction and will be tested in 2010.

\subsection{GRPC detector development}

GRPCs are well tested detectors. They have been successfully used in BELLE 
for almost a decade and more recently in OPERA and ALICE. Nonetheless, 
GRPCs to be used as the sensitive medium in the future HCAL need to be thinner 
than the standard ones in order to reduce the total radius of the hadronic 
calorimeter and hence the cost of the magnet coil inside which the HCAL 
should be placed.  For this reason, GRPCs of 3\,mm thickness were conceived 
and built. 
In addition to the thickness constraint, dead zones due to spacers used 
between the two glass plates of the GRPC are to be reduced. New schemes 
using tiny ceramic balls were introduced. This reduces the dead zones 
from a few percent to a negligible level. Another important point is the gas 
consumption in such a detector. New gas distribution designs were studied 
with the goal of renewing the gas efficiently, increasing the homogeneity and 
at the same time reducing the needed gas flow.    Another R\&D activity 
concerning the resistive coating of GRPCs is going on. Different coatings 
were tested to reduce the number of pads associated to one mip.  This 
number was found to go from 1.6/mip in case of standard graphite coating 
to 1.3/mip for more resistive products like the Licron and Statguard 
products. To guarantee the homogeneity of the painting on detectors of 
large surface area silk screen printing techniques were successfully used. 
Stability of high voltage connections were also improved by using new 
gluing materials.  Although the accumulated charge on the HCAL GRPCs in 
the future ILC experiments is expected to be very small, a few GRPCs are 
being exposed to high irradiation in the GIF facility at CERN to discover 
any eventual ageing effect.

A recent development has started recently to improve on the detection rate 
capacity. Due to the float glass resistivity ($\sim10^{13}\Omega$.cm) 
standard GRPCs can 
operate efficiently only with rates below 100\,Hz/cm$^2$.  New semi-conductive 
glass ($\sim10^{10}\Omega$.cm) developed in China by the Tsinghua group 
was used to build 
small GRPCs. Exposure to pion beams at CERN has shown that the new GRPC can 
operate up to 30\,kHz/cm$^2$ with the same efficiency. This breakthrough in 
GRPC technique is an important achievement not only for future ILC 
experiments but also for ones like SLHC where very high rates are 
expected.

Using cosmic ray test benches or test beams at CERN, the efficiency of the 
different  GRPCs was found to be more than 90\% when operated in the high 
voltage range between 7 and 8\,kV.

\subsection{Micromegas detector development}

The impressive progress made in the last decade on Micro Pattern Gaseous Detectors
makes Micromegas a viable choice for a semi-digital hadronic calorimeter. 
This detector is used in several physics experiment: it is part of the COMPASS tracker, 
the T2K TPCs and is also a serious candidate for the upgrade of the ATLAS muon spectrometer 
at SLHC and for an ILC TPC.

Micromegas is a proportional gaseous detector and is therefore best suited for 
the application in a semi digital HCAL, the relation between readout signals 
and primary ionisation being strictly linear. 
It works in standard mixtures like Ar/\textit{i}C$_{4}$H$_{10}$ and Ar/CO$_{2}$ and at voltages below 500~V. 
Thanks to its mild amplification field (40~kV/cm) and the small size of the 
amplification gap (128~$\mu$m) the rate capability of Micromegas is very high and 
its gas gain remains stable after a total accumulated charge as high as several mC/mm$^{2}$. 
It is hence well suited for an experiment that should run for several decades.

With the Bulk fabrication technique developed and routinely used by the CERN technical service, 
large area and robust Micromegas can be made. To reduce the detector thickness to a minimum, 
the manufacturing process was modified to integrate on both side of a single PCB, 
the readout (digital front-end chips) and the amplification part (pads and mesh). 
With analog electronics, a detection efficiency to 100~GeV/c muons of 98~\% was measured with 
a mean number of hits per traversing particle below 1.15. Micromegas thus satisfies the SDHCAL 
requirements of thickness, efficiency and pad multiplicity.

As all MPGD, Micromegas has a tendency to spark when operated at high gas gains. 
For that reason, additional passive components are connected with the front-end chips to the PCB 
and have shown to fully protect the electronics. 
Other implementations like buried resistors and capacitors and new protective methods 
and schemes are currently under investigation.

\subsection{Electronics}
To read out the gaseous detectors mentioned above an electronic chip called 
HARDROC with a semi-digital readout was developed and successfully tested. 
The chip has three thresholds (2-bit readout).  It has 64 channels and 
each of the 64 channels is made of:
\begin{itemize}
\item	Fast low impedance preamplifier with a variable gain over 6 bits 
per channel
\item	Variable shaper (50-150ns) and Track and Hold to provide a 
multiplexed analogue charge output up to 15pC. 
\item	3 variable gain fast shapers (15ns) followed by 3 low offset 
discriminators which allow handling wide dynamic range from 10fC up to 
15pC. The thresholds are loaded by three internal 10-bit  DACs. 
\end{itemize}
In addition, the chip has a 128 deep digital memory to store the 
discriminators' outputs and the bunch crossing identification coded over 24 
bits counter. It is equipped with a power pulsing system which allows  
a consumption lower than 10$\mu$W/channel to be reached with a 0.5\% 
duty cycle. The 
cross-talk among the 64 channels was measured and found to be less than 
2\%.

\vspace{0.2cm}
\noindent
In parallel to the development of HARDROC, another 64 channel ASIC called DIRAC was fabricated. 
It is intended for equipping both RPC and Micromegas and accordingly has two dynamic ranges: up to 10~pC or 200~fC. 
It has already been successfully tested in the latter detector.
Each channel is equipped with a switched charge preamplifier, a switched baseline restorer and three comparators. 
The threshold of each comparator is set by an 8-bit DAC. The 2-bit result of the comparison is stored into an 8 
event depth memory. Moreover, each event is stamped with a 12-bit identifier for bunch crossing identification. 
Configuration and readout signals are serial digital signals, this allow several ASICs to be chained in view of 
the construction of large area detectors.

The DIRAC is auto-triggered and designed so as to minimize the threshold dispersion over all channels and 
avoid any calibration. The measured dispersion is better than 1.17~fC (in the Micromegas mode) which is small 
compared to the typical MIP input charge of 25~fC. 
Its functioning is synchronous to a clock:  signals are detected during the first half period of the clock 
while threshold comparison and memory writing occur in the second half period. This functioning together with 
power pulsing capability (10~$\mu$W/channel) are intended to match the ILC beam time structure.

\subsection{Small prototypes}
 
To validate the concept of a semi-digital hadronic calorimeter, a Printed
Circuit Board was developed to host 4 HARDROCs. The board provides the 
connection between adjacent chips as well as 
linking the first chip to the readout 
system. For simplification reasons the readout system using FPGA devices 
was also implemented on the same PCB as well as a USB device responsible 
of the communication between the FPGA and an external server. 
The PCB is an 8-layer , 800~$\mu$m thick circuit. On one of the two PCB 
faces, 256 copper pads of 1$\times$1\,cm$^2$ were printed. 
The distance between two 
adjacent pads was chosen to be 500~$\mu$m. The cross-talk among adjacent 
pads was tested before other electronics components were fixed on the PCB 
by injecting a charge of 1\,pC on one pad using an appropriate probe. The 
charges induced on the adjacent pads were then measured and found to be 
less than 0.3 \%.  Acquisition software was also developed. It permits  
the configuration parameters to be downloaded to the different chips and 
data to be collected from these chips through the FPGA device. 
Two readout modes 
were implemented. The first one is an ILC-Like one where events are 
recorded during the bunch crossings and the readout takes place after. 
The other mode was conceived for cosmic rays and beam test studies. In 
this mode the acquisition and data taking is stopped when an external 
trigger occurs. The memory of the different chips is then read out. In 
both modes each event is associated with a time stamp. In the external 
trigger mode the time difference between the external trigger and the last 
recorded event is also given. This determines the time occurrence of each 
event with respect to the external trigger one.  The time precision is 
given by the HARDROC internal clock which runs at 5\,MHz frequency.

The electronics boards described above were attached to small GRPCs and 
assembled together to form a small prototype of five chambers. The 
acquisition system was extended to deal with the data coming from each 
chamber and assemble them in events. 
A LABVIEW-based graphic interface was developed. This allowed an easy gain 
correction of all the channels. 
The setup was exposed first to cosmic rays and then to beams at CERN. The 
whole system performed very well and allowed the GRPC efficiency 
and multiplicity to be studied. 
Completed with 2\,cm stainless steel plates, the setup 
was exposed to pions at the CERN PS to study the first phase 
of hadronic showers.

\vspace{0.2cm}
\noindent
Several Micromegas prototypes equipped with analog or digital electronics 
and sizes between 8$\times$8~cm$^{2}$ up to 32$\times$48~cm$^{2}$ were fabricated and tested in a beam at CERN SPS and PS.
Prototypes with analog electronics were used for extensive characterisation in 2008. 
Signal distribution from high energy particles, uniformity, efficiency, pad multiplicity were measured. 
Their dependence on pressure, temperature and threshold variations are now fully understood. 
This work was summarized in a publication in JINST. 

Tests of Micromegas chambers equipped with 1 DIRAC or 4 HARDROCs or 24
HARDROCs chips and a Detector Interface Board (so-called DIF) were performed in 2009. 
The data acquisition chain consists of the ASICs, the DIF, a USB cable and
either the CrossDaq or LabView based software.
The latter was developed for chip characterisation and recently for more flexible and efficient data taking.
Multiplicity measurements are compatible with previous values from
analog readout prototypes.
The analysis of the collected data is on-going.

\subsection{Technical prototypes}

The success of the small prototype was the first step towards the 
construction of the technical prototype. The second and decisive step 
is to build a fully equipped detector of 1 m$^2$. For this purpose, a new 
PCB hosting 24 ASICs was designed with the possibility to connect a few of 
them together. An independent interface board (DIF) connecting the PCB to the 
acquisition system was also produced and tested. Six such PCBs were 
produced and equipped.  Every two PCBs were connected to each other and 
connected to one DIF.  The six boards were fixed to a mechanical support 
made of stainless steel plate and then attached to a 1~m$^2$ GRPC. The three 
DIFs connected to the PCBs are chained together and connected to a 
monitoring computer.  An acquisition system using the Xdaq system 
developed  by the CMS collaboration is used to build  events from the 
collected data.  A cosmic ray test bench was used to study the whole 
system. After a debugging period, the whole system works adequately. Few 
variants of GRPC were tested using the large electronic board in test beam 
at CERN in June and August 2009. The collected data were analysed. 
Synchronization among the different DIFs was successful and the data 
coherence demonstrated.  The measured efficiency of the 1\,m$^2$  GRPC was 
found to be in the same range of those of small GRPC chambers($>90$\%). This 
success constitutes an important milestone in the validation of the 
SDHCAL. To complete this study an improved electronic board equipped with 
the new version of the HARDROC chips is under construction. The board will 
be attached to a 1~m$^2$ GRPC and inserted into a cassette.  This will be the 
first unit of the technical type to be built in 2010.

\vspace{0.2cm}
\noindent
The 1~m$^{2}$ Micromegas prototype is formed by 6 PCBs of
32$\times$48~cm$^{2}$, each equipped with a mesh and 24 HARDROCs.
These units are called Active Sensor Unit (ASU) and are placed into
the same gas chamber with minimum dead areas between the ASUs. Our
choice was to lay one mesh per ASU instead of a 1~m$^{2}$ mesh on the
6 ASUs, essentially to avoid destructive sparks from the increased
capacity of a too large area mesh. Moreover no facility can yet
perform the complete bulk construction procedure over a 1~m$^{2}$ area.

Before proceeding to the construction of a technical prototype, a
mechanical prototype without electronics and mesh was assembled to
check the gas tightness and validate the design and the complete assembly procedure.
The technical prototype is now being assembled and will be tested in 2010 at CERN. 
The commissioning of the 6 ASUs has started a few
months ago already. It consists of HV training in air and in gas, ASIC precise
calibration, electronic chaining of the ASUs by pair, measurement of
the response to $^{55}$Fe quanta and cosmic particles in a gas box.
These steps are essential to insure the overall quality of the
1~m$^{2}$ prototype and are now routinely performed in the lab. This
same procedure will be also used for the fabrication of additional
prototypes, based on HARDROC or DIRAC chips, in 2010.

The fine granularity of Micromegas chambers (1~cm$^{2}$), its
insensitivity to neutrons and its fast response are particularly
interesting for a W-based HCAL for a multi-TeV range linear collider.
Indeed, the shower separation in jets at CLIC will be more difficult
than at ILC and a precise timestamping crucial.
In view of a possible application in a CLIC HCAL, it is foreseen to
equip a large scale tungsten structure with Micromegas layers.

\subsection{Preparation for the 1~m$^{3}$ technical prototype}

The technical prototype will be made of 40 detectors interleaved with 
2\,cm stainless steel plates. The mechanical structure of the prototype is 
being currently designed.  The aim is to have a self-supporting structure 
like the one proposed in the ILD and SiD concepts. The detectors with 
their readout boards will be assembled into cassettes that will be inserted 
between the Stainless Steel plates. 
The readout system will include data concentrators which will connect the 
DIF mentioned above to the general CALICE acquisition system. The 
development of data concentrators is almost finished and the whole readout 
system will be tested very soon to validate the whole chain.  A gas 
distribution system controlling both the gas flow and pressure in the 
different chambers is under construction in collaboration with the CERN 
Gas Service. It will provide gas mixtures for RPCs (TFE/\textit{i}C$_{4}$H$_{10}$/SF$_{6}$) 
as well as for Micromegas (Ar/\textit{i}C$_{4}$H$_{10}$ or Ar/CO$_{2}$) chambers.
High voltage power supplies using the Cockroft-Watson were 
designed and will be produced very soon. In the case of Micromegas which 
is works at voltages below 500~V, a CAEN crate with 40 HV channels is already available. 
The construction of the technical prototype is expected to start at the end of 2009 and to be 
completed by 2010. In addition to the hardware development, software 
activity is going on in order to prepare properly the comparison between 
data and the hadronic shower models used in the simulation.

\section{Technical prototype front end electronics}
\label{sec:FEE}
A second generation of readout ASICs have been developed to read out
the technical prototypes defined in EUDET and CALICE.  These are based on
the first generation of chips that were used for the physics prototype for the
analogue front-end part but add several essential features:
\begin{itemize}
\item	Auto-trigger to reduce the data volume
\item	Internal digitization to have only digital data outputs
\item	Integrated readout sequence to minimize the number of  lines
between chips
\item	Power-pulsing to reduce the power dissipation by a factor 100
\end{itemize}
Two chips have been designed, following the EUDET milestones:
\begin{itemize}
\item	HARDROC for digital hadronic calorimeters, using RPCs or
micromegas chambers
\item	SPIROC for analogue hadronic calorimeters and also for Si-W
electromagnetic calorimeters.
\end{itemize}
Both chips have been successfully tested on testbench so that they are
ready for tests with detectors. A third chio, SKIROC, has been partially 
developped, the analogue part, optimized for the ECAL, has been successfully 
tested earlier. A second version, with very large dynamic range and 
64 channels, is underway. 

Altogether these chips are being developped in an ``industrialized'' manner,
with many building blocks in common, such that one design benefits from the 
experience with the other. The digital part was first developped and 
optimized with the HARDROC chip which is the simplest from the analogue
point of view. SPIROC has a switch for the preamplifier characteristics 
which allows to toggle between usage for SiPM or silicon diode pad readout
and is being uesed to proceed with the development of the Si-W ECAL 
technological prototype until the final SKIROC version is availabe.  

\subsection{DHCAL technical prototype: HARDROC ASIC}

\begin{figure}\centering    
\includegraphics[angle=0,width=0.6\textwidth]{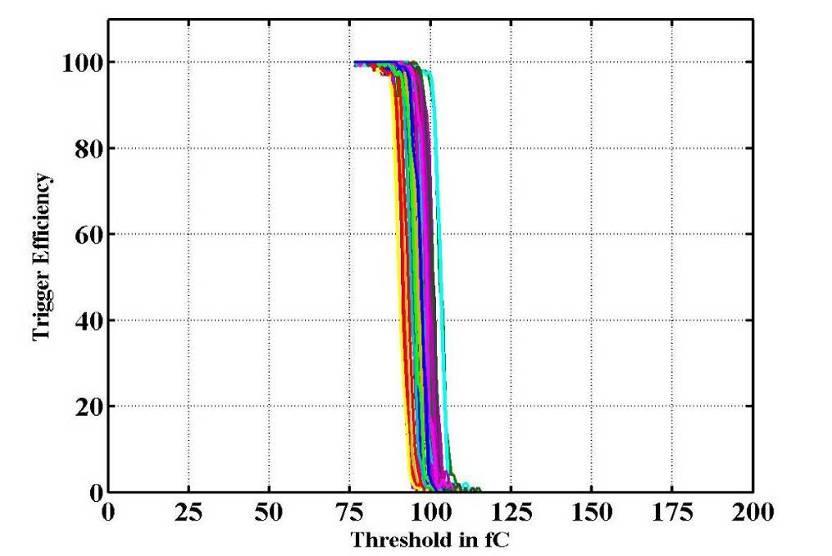}
\caption[]{\label{fig_chris_1}\em Trigger efficiency for 100\,fC input as a function of DAC
threshold.}
\end{figure}

HARDROC readout is a semi-digital readout with three thresholds (2 bits
readout) which allows both good tracking and coarse energy measurement,
and also integrates on chip data storage. The chip integrates 64 channels
of fast, low impedance current preamplifier with 6-bit variable gain
(tuneable between 0 and 2\,V), followed by a fast shaper (15\,ns) and low
offset discriminators. The discriminators feed a 128-deep digital memory
to store the 2$\times$64 discriminator outputs 
and bunch crossing identification
coded by a 24-bit counter. Each is then read out sequentially during the
readout period.
A first version has been fabricated in AMS SiGe 0.35\,$\mu$m technology in
September 2006 and met design specifications. A second version was
produced in June 2008 to fit in a smaller low-height package (TWFP160)
which necessitated changing the double-row bonding pad ring into a single
row, rerouting all the inputs and removing many many pads. A possibility
for a third threshold was added at that time, also separating more widely
the three thresholds (typically 0.1--1--10\,pC) and the ``power off''
state dissipation was brought down to a few $\mu$W for the whole chip.
The chip performance is shown in Fig.~\ref{fig_chris_1}. 
The trigger efficiency allows the MIPs for RPCs 
to be discriminated with 100\,fC threshold (10\,fC for micromegas),
with a noise of 1\,fC. The power pulsing scheme has also been validated,
also shown in Fig.~\ref{fig_chris_1}, where 25 $\mu$s 
are required to start up the chip so
that it can trigger on a 10\,fC input signal. Finally the readout scheme,
which is common to all the chips has been valicated on the large
square-meter board, built as a scalable technical prototype of the SDHCAL,
that is readout from the side.

\begin{figure}\centering    
\includegraphics[angle=0,width=0.6\textwidth]{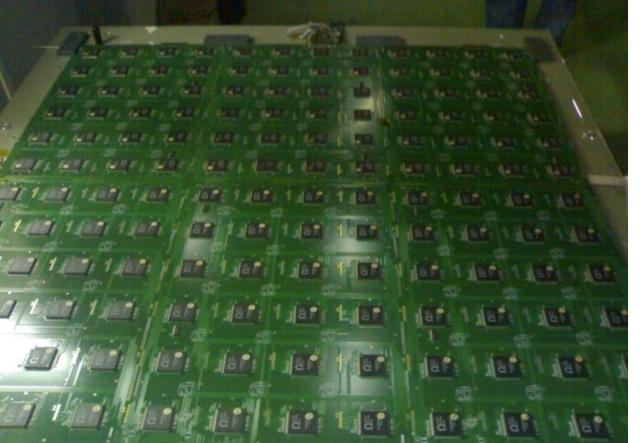}
\caption[]{\label{fig_chris_2}\em Layout of HARDROC2, view of the chip packaged in TQFP160 and square meter prototype of RPC DHCAL with 144 HARDROC.}
\end{figure}

\subsection{AHCAL technical prototype: SPIROC ASIC}

\begin{figure}\centering    
\includegraphics[angle=0,width=0.6\textwidth]{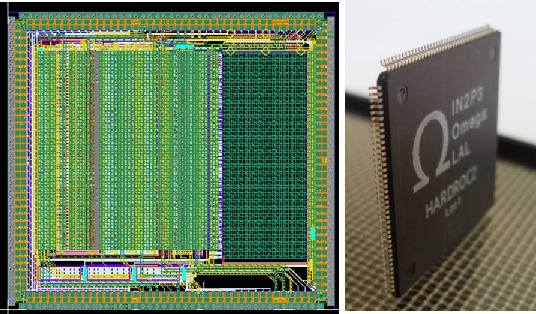}
\caption[]{\label{fig_chris_3}\em Layout of SPIROC1.}
\end{figure}

\begin{figure}\centering    
\includegraphics[angle=0,width=0.6\textwidth]{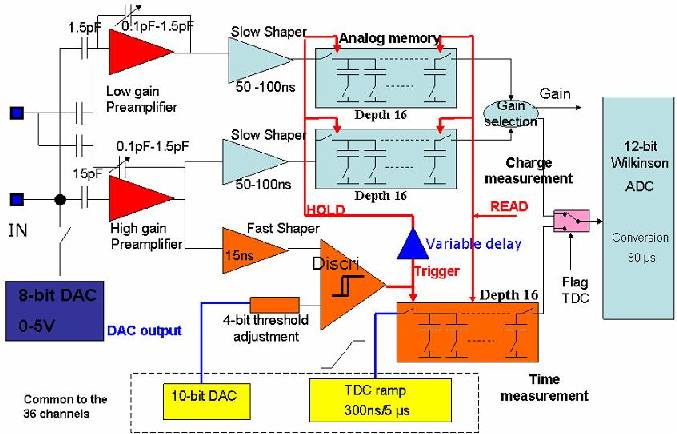}
\caption[]{\label{fig_chris_4}\em Schematic diagram of one channel of SPIROC1.}
\end{figure}

The SPIROC ASIC, which reads 36 SiPMs, is an evolution of the FLC\_SiPM used
in the AHCAL physics prototype. As for the ECAL, it keeps most of the analogue
part, adding an analogue memory to record up to 16 events of a train and
the auto-triggering capability. The digitization is implemented inside the
chip as well as the data handling.
The first prototype was fabricated in June 2007 in AMS SiGe 0.35\,$\mu$m.
The chip layout is shown in Fig.~\ref{fig_chris_3}. 
Its area is 35\,mm$^2$ for 36 channels
and it is packaged in a CQFP240 package. Similarly to HARDROC, a second
version has been realized in June 2008 to accommodate a thinner TQFP208
package and fix a bug in the ADC.
The schematic diagram of one channel is shown in Fig.~\ref{fig_chris_4}.

\begin{figure}\centering    
\includegraphics[angle=0,width=0.6\textwidth]{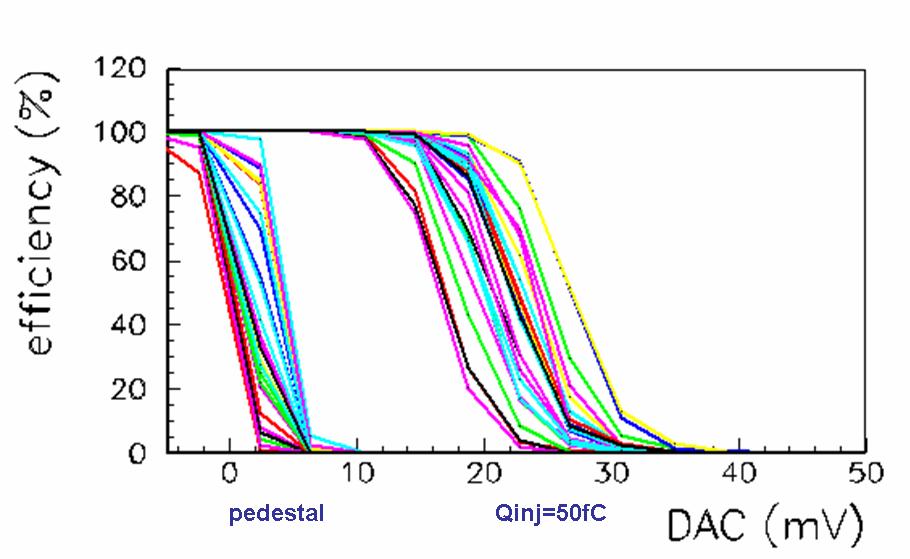}
\caption[]{\label{fig_chris_5}\em Efficiency of auto-trigger mode. An injected signal of 50 fC is separated from pedestal with a S/N ratio of 8.}
\end{figure}

The main features are:
\begin{itemize}
\item	The 8-bit input DAC has been conserved, although its power
dissipation has been brought down by 3 orders of magnitude to 
1\,$\mu$W/channel
as it is not power pulsed. The DAC also has the particularity of being
powered with 5V whereas the rest of the chip is powered with 3.5V.
\item	The voltage amplifier architecture with variable gain has also
been kept, with a gain variable on 4 bits. However, the high gain/low gain
separation is now done at the preamplifier level by having two preamplifiers
in parallel on the input.
\item	The charge is measured on both gains by a ``slow'' shaper (50--150\,ns)
followed by an analogue memory with a depth of 16 capacitors.
\item	The auto-trigger is taken on the high gain path with a high-gain
fast shaper followed by a low offset discriminator. All these blocks are
new. The discriminator output is used to generate the hold on the 36
channels. The threshold is common to the 36 channels, given by a 10-bit
DAC similar to the one from HARDROC with a subsequent 4-bit fine adjustment per
channel.
\item	The discriminator output is also used to store the value of a
300~ns ramp in a dedicated analogue memory to provide time information with
an accuracy of 1ns
\item	A 12-bit Wilkinson ADC is used to digitize the data at the end of
the acquisition period.
\end{itemize}
The digital part is complex, as it must handle the write and read
pointers, the ADC conversion, the data storage in a RAM and the readout
process.
The chip has been extensively tested since October 2007. The first series
of tests has been mostly devoted to characterizing the analogue performance,
which meets the design specifications. A single photoelectron spectrum
displayed in Fig.~\ref{fig_chris_5} 
shows  a photo-electron to noise ratio around 8. The
auto-trigger is also shown in Fig.~\ref{fig_chris_5} and allows 
a threshold down to 50 fC to be tested.

The digitization part has also been characterized and the 12-bit ADC
exhibits a very good integral non-linearity of 1 LSB and a noise
between 0.5 and 1 LSB.

The chips are now being assembled on an AHCAL PCB (HBU) and tested with
the detector, as described in Sec.~\ref{sec:AHCAL}.

A new effort has been started to redesign the analogue input stage 
and further optimize it for SiPM readout. According to siumulations, 
it improves the signal-to-noise ratio by a factor 3 - 4. Once successfully
tested with analogue chips, it is foreseen to integrate the design 
into the SPIROC architecure.   

\section{Technical prototype DAQ}
\label{DAQsection}
The second generation data acquisition (DAQ) system follows an approach of 
using commercially-available hardware, where possible, with less 
dependency on 
bespoke equipment than has traditionally been the case
in high energy physics and 
specifically in the previous CALICE DAQ system.  The system is designed to 
be 
generic and hence applicable to many sub-detector systems and also to be able 
to provide the readout for the next generation of CALICE/EUDET prototype 
calorimeters.  Indeed the CALICE calorimeters are an ideal test-bed as 
there 
are the different types (electromagnetic, hadronic, digital, etc.) with 
additional instrumentation also included during test-beam running such as 
small 
tracking systems.  In principle, the DAQ system should be capable of 
reading out 
the final ILC calorimeter, with potential upgrades in chip flavor and 
network 
speed, but maintaining the basic architecture.  Therefore the system 
should be easily 
upgradable, both in terms of ease of acquiring new components and 
competitive 
prices. 

\subsection{Overview of the DAQ system}
An overview of the DAQ system is shown in Fig.~\ref{fig:daq-overview}; 
an explanation of each component is given below:
\begin{itemize}

\item A control PC has the DAQ software (currently DOOCS~\cite{doocs}, 
developed for 
      XFEL) and will control the various DAQ PCs.

\item A customised DAQ PC is made from standard components but optimised 
for 
      having two off-detector receiver (ODR) cards and also the large 
storage 
      facilities which will be required.

\item The ODR is a commercial development board~\cite{plda} with a large 
FPGA, 
      optical fibers and a PCI Express bus.  This can have a maximum of 
four 
      optical fibers going to the next stage upstream.

\item The four optical fibers from the ODR will connect to the link data 
aggregator 
      (LDA) which is a commercial FPGA development board~\cite{enterpoint} 
with add-on 
      ethernet (to 
      connect to the ODR) and HDMI (to connect upstream) boards.  This is 
essentially 
      a concentrator card which takes data from upstream on 10 HDMI links 
and merges 
      them into one ethernet link to the ODR.

\item At the end of the calorimeter slabs, a detector interface (DIF) card 
is an 
      electronics board which is specific to each calorimeter but converts 
the data 
      into a common structure and format so that it is sent downstream, 
off the detector,  
      in a uniform way.  Each of the three DIFs in the present system
(AHCAL, DHCAL and ECAL) is custom 
      designed.

\item Providing the interface to the accelerator or a standalone clock is 
the clock 
      and control card (CCC).  This card was designed in-house due to the 
problems 
      of finding commercially available systems which time-in several 
components all 
      running with different clocks.

\end{itemize}

~\begin{figure}[htp]
\begin{center}
\includegraphics[height=6.5cm]{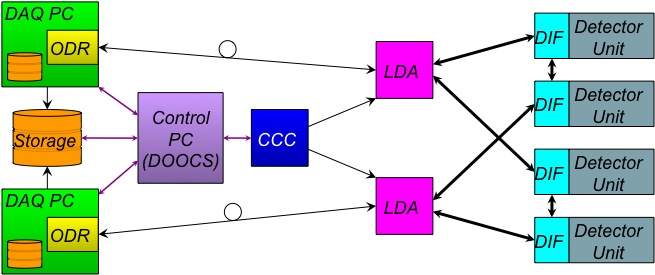}
\caption{\label{fig:daq-overview}\em Architecture of CALICE DAQ system.}
\end{center}
\end{figure}

\subsection{Status of the components}
Overall, all electronics cards have been prototyped, where appropriate, 
and all 
are now in, or have finished, the production phase.  Hardware orders are in 
place to ensure that we
have enough systems available for laboratory, beam tests and spares.  
Significant blocks 
of firmware have been produced and final debugging of the whole DAQ chain 
is being 
performed.  As well as the full chain which is being debugged in the UK, a 
complete 
system has been delivered to LLR to enable further debugging and 
integration 
with the detectors. The status of each component is discussed briefly 
below.
\begin{itemize}
\item
\underline{DIF}:
The ECAL DIF has been designed by UK groups and is now in production.  
This 
credit-card-sized board has been somewhat miniaturised from the prototype 
and the two production cards
already manufactured are in use for the full system tests.  To cover the 
expected 
need of 30 DIFs, 40 will be manufactured in the coming months.  The AHCAL 
DIF and DHCAL 
DIF are the responsibility of the detector groups, primarily DESY and the 
French 
groups and are discussed in Sec.~\ref{sec:AHCAL} 
and~\ref{sec:sDHCAL}.
\item
\underline{LDA}:
Final debugging of the DIF$\Rightarrow$LDA link is ongoing.  To cover the 
estimated 
10 LDAs needed in test-beam running, 20 baseboards have been procured and 
20 of 
each of the ethernet and HDMI add-on boards are on order.
\item
\underline{CCC}:
At most, three CCCs will be needed at any one time; ten have been 
manufactured and 
tested and are ready for use.  The one outstanding issue is the link from 
the CCC to 
the LDA.  A small add-on PCB has been designed and is under test.  
Production will 
proceed after that and we expect to do this in-house as the 
design is 
so simple.
\item
\underline{ODR and DAQ PC}:
As the ODR was the first component bought and the basis of the DAQ system, 
this has 
been stable for some time.  Eight ODRs and six DAQ PCs have been purchased 
to provide 
for the expected need of four and three, respectively.
\item
\underline{DAQ software}:
After an initial survey of available software by UK groups, it was decided to use the 
DOOCS system, 
developed by XFEL.  Since responsibility of the software has been passed 
to LLR, 
this is being reconsidered, particularly in light of the possibility of 
combined test 
beams and as part of a common DAQ effort in the EU FP7 bid, AIDA.
\end{itemize}
The major issues still to be done for the DAQ system are:
\begin{itemize}

\item Debugging the DIF$\Rightarrow$LDA link which is the only link in the 
full system 
      yet to be established.

\item Production of all final components, specifically manufacturing the 
DIFs and 
      receipt of the LDA add-on boards from the manufacturer.

\item Finalise simple add-on PCB for CCC and LDA link.

\item Full documentation of components on the CALICE {\tt twiki} (already 
started).

\item Archiving of all code in a central {\tt svn} code repository 
(already started).

\end{itemize}

It is expected to achieve all of the above and hence have a working DAQ 
system 
with a full set of components (plus spares) by the end of the year.  This 
will then 
allow the LLR group to set-up their system and integrate with the detector 
as 
well as then providing other detector groups with a system and spares.

\section{Future test beam plans}

\subsection{Physics prototypes}
The years 2010-11 will see the finalisation of the main physics prototype phase. 
As described above, a physics prototype of 
the DHCAL based on RPCs will be completed in 
the first half of 2010. As for previous testbeams including the
AHCAL, there will be data taking in combination with 
the Si-W ECAL and the TCMT, as well as standalone DHCAL data taking. Including 
commissioning and calibration phases altogether 14 weeks of 
test beam time will be requested from FNAL. Within this 14 weeks, CALICE 
should be the primary beam user for about 8 weeks. The other 6 weeks 
can be spent with parasitic running or the setup of the experiment. 
The physics program to be conducted will be largely oriented on that which 
has been conducted in the corresponding data taking in the years 
2006-09. 
In the combined running, the emphasis will be put on energy ranges in 
which it is expected to see signals in the electromagnetic part and the 
hadronic part 
(plus tail catcher) In the standalone running also low energy hadrons are 
to be collected as well as electrons. 
Priorities would have to be defined later on but the data which 
were already taken give good guidelines.
It is also envisaged to replace a few layers of the DHCAL
with GEMs as sensitive detectors.
This may happen towards the beginning of 2011. This 
effort might face the constraint that due to customs regulations, 
the CALICE stage currently at FNAL is required to be 
shipped back to Europe in April 2011; in that case tests can be completed at CERN.

The one remaining detector which is yet to complete a physics
prototype detector is the DECAL; this is expected to be achieved
by 2012.

A new initiative has been started in order to study the properties of tungsten 
as aborber material, primarily for an HCAL at a multi-TeV collider, but possibly 
also for the ILC. 
In the versatile structure of the physics prototype the steel absorber plates
can be replaced by tungsten plates. Tests with existing scintillator layers
could start in 2011, tests with gaseous layers as they become available.  

\subsection{Technical prototypes}
The previous sections indicate that the CALICE collaboration is entering a 
new phase of R\&D in which readout 
technologies and mechanical designs do meet already many requirements of 
the operation in a detector for a Linear Collider. 
Several groups of the collaboration are already quite advanced and new 
full scale prototypes are expected towards the end of 2010. 
The finalisation of these prototypes will be preceded by a number of 
larger and smaller testbeam efforts which will allow for 
maturing the newly developed technologies.
Examples for these test beam efforts are:
\begin{itemize} 
\item Testbeams with 1\,${\rm m^2}$ units of the technical prototype 
of the SDHCAL (both GRPC and micromegas variants). These 
units might already
be part of the production of the entire prototype. 
\item For the analogue hadron calorimeter it is envisaged to have a 
smaller scale testbeam in 2009 to prepare for electronics commissioning 
followed by a so-called horizontal test in 2010 and a vertical test 
in 2011.  This means the available 
equipment will be arranged to allow for the measurement of electromagnetic 
showers.
\item The Si-W ECAL is planning to make tests with single ASU towards the 
end of 2010 in an electron testbeam.
\end{itemize}

It has to be stressed that the primary goal of these prototypes is to 
study technological solutions for the  calorimetry at the ILC. 
The strategy for the coming years should take this into account. Here the 
main keywords are power pulsing and the limited depth of the 
buffers in the VFE which allow only for data taking rates of the order of 
100\,Hz during a spill. In addition to the pure technological 
issues a physics program is to be pursued. This physics program is  
derived from those of the physics prototypes, taking the new technical 
constraints into account. It requires the operators of 
testbeam sites to actively respond to the needs of the CALICE 
(LC) testbeam data taking at an very early stage. As it is foreseeable 
that potential high statistics physics runs will take a 
considerable amount of time, and this will require the deployment of remote 
control at the experimental sites.

A first large scale testbeam with a fully equipped technical prototype 
of an SDHCAL can be expected towards spring 2011. It is still to be 
clarified
in what proportion this SDCHAL will be equipped with the two technologies 
under study, namely using GRPCs or micromegas as sensitive devices. The 
CALICE collaboration will identify this on the basis of experience gained 
with the two technologies by laboratory studies and during test beam 
campaigns of the year 2009.

Ideally, the SDHCAL will be joined by an Si-W ECAL technical prototype 
by end of 2011. The running of an AHCAL technical prototype 
alone and together with the Si-W ECAL technical prototype is to follow. 
During the year 2010 mechanical interfaces between the 
different detector 
types will have to be defined. More general year 
2010 is to be used to integrate the detector components with the newly 
developed DAQ systems in order to provide an efficient data taking. 

The program which is summarised in Table~\ref{tab:testbeams} requires a 
high availability of testbeam areas The CALICE management and the CALICE 
TB together with the corresponding 
ILC R\&D panels will work out until December 2009 whether ILC detector 
R\&D can occupy consecutively testbeam areas for a time of two or more 
years starting with the beginning of 2011. Such a high availability of 
testbeam areas would also allow for an easier conduction of smaller 
testbeam
programs as for example with the DECAL, which has had a somewhat uncertain 
time scale due to funding issues in the UK. In addition a well 
functional infrastructure would facilitate the testing of a prototype for 
the electromagnetic calorimeter based on scintillating tiles (Sc-W ECAL) 
of which one layer
can be expected towards the end of 2012. 
Finally, technological prototype layers with timing capabilities should also 
be used in a beam test with a tungsten absorber structure. 

\begin{table}[htdp]
\begin{center}
\begin{footnotesize}
\begin{tabular}{@{} |ccccccc| @{}}
 \hline
    Project & 2010/1 & 2010/2& 2011/1 & 2011/2 & 2012/1 & 2012/2 \\
    \hline
    Phys. Prot. Si-W ECAL/DCHAL/TCMT& xx & xx & xx & - & - & -\\
    Phys. Prot. W ECAL / W HCAL / TCMT&  & x & x & xx & xx  & -\\
    Tech. Prot. DHCAL & x & x & xx & xx & xx & xx\\
    Tech. Prot. AHCAL & x & x & x & x & xx & xx\\
    Tech. Prot. Si-W ECAL & - & x & x & xx & xx & xx\\
    Phys. Prot. DECAL & x & x & x & x & x & x\\
    Tech. Prot. Sc-W ECAL & - & - & - & - & - & x\\
    \hline
\end{tabular}
\end{footnotesize}
\end{center}
\caption{
\label{tab:testbeams}
\em The table indicate the envisaged testbeam activities until the 
end of 2012. The symbol {\bf --} means ``No activity planned'', The symbol 
{\bf x} means ``Test of small units can be expected'', The symbol {\bf xx} 
means ``Large scale testbeam planned''.}
\end{table}%

As a concluding remark it needs to be stressed that the time plans 
presented here depend essentially on the available funding to complete the
various prototypes. Thus, the actual running of, in particular, the test 
beams with fully equipped prototypes could easily be delayed by one year 
with respect to the dates indicated in Table~\ref{tab:testbeams}.

\section{Conclusions and request}

CALICE has successfully completed a first round of data taking at CERN and
Fermilab with physics prototypes for silicon based ECAL and scintillator
based ECAL and HCAL. The detectors show the expected performance and
the test beam data start to provide constraints in hadronic shower models.
The beam test of a gaseous digital HCAL prototype is under preparation and
supposed to start mid 2010. The physics program will the be extended
towards the study of tungsten as absorber material.

Various technological prototypes are under preparation and in different
stages of commissioning or beam test with detector elements. A program
has been set up to demonstrate the viability of the designs with larger 
structures
exposed to beams. Novel technologies like the digital ECAL have also
undergone first tests with encouraging results.

We are asking the PRC for endorsement of this program and in particular
its scope of technologies and its timeliness with respect to the ILC 
schedule.

We request the continuation of DESY's support for test beam campaigns
at Fermilab and CERN, for the adaptation of the set-up to new devices, for
the continuation of electronics integration development towards completion
of technological demonstrator prototypes, and for the continued provision
of computing resources.

\section{ Acknowledgments}

We would like to thank the technicians and the engineers who
contributed to the design and construction of the prototypes.
We also gratefully acknowledge the DESY, CERN and FNAL managements for their support and
hospitality, and their accelerator staff for the reliable and efficient
beam operation. 

We would like to thank the HEP group of the University of
Tsukuba for the loan of drift chambers for the DESY test beam.
The authors would like to thank the RIMST (Zelenograd) group for their
help and sensors manufacturing.
This work was supported by the 
Bundesministerium f\"{u}r Bildung und Forschung, Germany;
by the  the DFG cluster of excellence `Origin and Structure of the Universe' of Germany ; 
by the Helmholtz-Nachwuchsgruppen grant VH-NG-206;
by the BMBF, grant no. 05HS6VH1;
by the Alexander von Humboldt Foundation (Research Award IV, RUS1066839 GSA);
by joint Helmholtz Foundation and RFBR grant HRJRG-002, Russian Agency for Atomic Energy, ISTC grant 3090;
by Russian Grants  SS-1329.2008.2 and RFBR0402/17307a
and by the Russian Ministry of Education and Science;
by MICINN and CPAN, Spain;
by CRI(MST) of MOST/KOSEF in Korea;
by the US Department of Energy and the US National Science
Foundation;
by the Ministry of Education, Youth and Sports of the Czech Republic
under the projects AV0 Z3407391, AV0 Z10100502, LC527  and LA09042  and by the
Grant Agency of the Czech Republic under the project 202/05/0653;  
and by the Science and Technology Facilities Council, UK.

\section{Appendix 1: The CALICE Collaboration}

\author{\centering
C.\,Adloff, 
J.\,Blaha, 
J.-J.\,Blaising, 
M.\,Chefdeville, 
C.\,Drancourt,
A.\,Espargliere, 
R.\,Gaglione, 
N.\,Geoffroy, 
Y.\,Karyotakis, 
J.\,Prast,
G.\,Vouters
\\ \it
Laboratoire d'Annecy-le-Vieux de Physique des Particules, Universit\'{e} de Savoie,
CNRS/IN2P3,
9 Chemin de Bellevue BP110, F-74941 Annecy-le-Vieux Cedex, France
}

\author{\centering
T.\,Cundiff, 
G.\,Drake, 
B.\,Haberichter, 
V.\,Guarino, 
A.\,Kreps, 
E.\,May, 
J.\,Repond, 
J.\,Schlereth, 
J.\,Smith\footnote{Also at University of Texas, Arlington},
D.\,Underwood, 
K.\,Wood, 
L.\,Xia, 
Q.\,Zhang
\\ \it
Argonne National Laboratory,
9700 S.\ Cass Avenue,
Argonne, IL 60439-4815,
USA}

\author{\centering
A.\,Brandt, H.\,Brown, K.\,De, C.\,Medina, J.\,Smith, J.\,Li, M.\,Sosebee, A.\,White, 
J.\,Yu
\\ \it
Department of Physics, SH108, University of Texas, Arlington, TX 76019, USA
}

\author{\centering
Z.\,Deng, Y.\,Li, Wang Yi, Q.\,Yue, Z.\,Yang
\\ \it
Tsinghua University, Department of Engineering Physics.Beijing, 100084, P.R.
China
}

\author{\centering
T.\,Buanes, G.\,Eigen, D.\,Fehlker, R.\,Roehrich, H.\,Sandaker
\\ \it
University of Bergen, Inst.\, of Physics, Allegaten 55, N-5007 Bergen, Norway
}

\author{\centering
Y.\,Mikami, 
O.\,Miller, 
N.\,K.\,Watson, 
J.\,A.\,Wilson
\\ \it
University of Birmingham,
School of Physics and Astronomy,
Edgbaston, Birmingham B15 2TT, UK
}

\author{\centering 
J.\,Butler, E.\,Hazen, S.\,Wu
\\ \it
Boston University, Department of Physics, 590 Commonwealth Ave.,
Boston, MA 02215, USA
}

\author{\centering 
M.\,J.\,Goodrick, 
T.\,Goto, 
L.\,B.\,A.\,Hommels, 
J.\,S.\,Marshall,
M.\,A.\,Thomson, 
D.\,R.\,Ward, 
\\ \it
University of Cambridge, Cavendish Laboratory, J J Thomson Avenue, CB3 0HE, UK
}

\author{\centering 
D.\,Benchekroun, 
A.\,Hoummada, 
Y.\,Khoulaki
\\ \it
Universit\'{e} Hassan II A\"{\i}n Chock, Facult\'{e} des sciences.\, B.P. 5366 Maarif, Casablanca, Morocco
}

\author{\centering 
J.\,Apostolakis, 
F.\,Duarte Ramos,
K.\,Elsener,
G.\,Folger, 
A.\,Gaddi,
H.\,Gerwig,
C.\,Grefe
W.\,Klempt,
L.\,Linssen,
D.\,Perini,
A.\,Ribon,
A.\,Sailer,
D.\,Schlatter,
P.\,Speckmayer,
V.\,Uzhinskiy
\\ \it 
CERN, 1211 Gen\`{e}ve 23, Switzerland
}

\author{\centering 
M.\,Oreglia
\\ \it
University of Chicago, Dept.\, of Physics, 5720 So. Ellis Ave., KPTC 201 Chicago, 
IL 60637-1434, USA
}

\author{\centering
M.\,Benyamna, 
N.\,Brun, 
C.\,C\^{a}rloganu,  
P.\,Gay, 
S.\,Manen, 
L.\,Royer
\\ \it
Laboratoire de Physique Corpusculaire de Clermont-Ferrand (LPC),
24 avenue des Landais,
63177 Aubi\`ere CEDEX, France
}

\author{\centering
J.\,Ha
\\ \it
Korea Atomic Energy Research Institute,
Taejon 305-600,
South Korea
}

\author{\centering
F.\,Abu-Ajamieh.
G.\,C.\,Blazey
D.\,Chakraborty
A.\,Dyshkant
K.\,Francis
D.\,Hedin
J.\,Hill
G.\,Lima
R.\,Salcido
V.\,Rykalin
V.\,Zutshi
\\ \it
NICADD, Northern  Illinois University,
Department of Physics,
DeKalb, IL 60115,
USA
}

\author{\centering 
V.\,Astakhov, V.\,A.\,Babkin, S.\,N.\,Bazylev, Yu.\,I.\,Fedotov, S.\,Golovatyuk, I.\,Golutvin, N.\,Gorbunov, 
A.\,Malakhov, S.\,Slepnev, I.\,Tyapkin, S.\,V.\,Volgin, Y.\,Zanevski, A.\,Zintchenko 
\\ \it
Joint Institute for Nuclear Research, Joliot-Curie 6,
141980, Dubna,
Moscow Region, Russia
}

\author{\centering 
D.\,Dzahini, 
L.\,Gallin-Martel, 
J.\,Giraud, 
D.\,Grondin, 
J.\,-Y.\,Hostachy, 
K.\,Krastev, 
L.\,Morin,
F-E.\,Rarbi
\\ \it
Laboratoire de Physique Subatomique et de Cosmologie - Universit\'{e} Joseph Fourier Grenoble 1 -
CNRS/IN2P3 - Institut Polytechnique de Grenoble,
53, rue des Martyrs,
38026 Grenoble CEDEX, France
}

\author{\centering 
N.\,D'Ascenzo, 
S.\,Christen,
U.\,Cornett, 
D.\,David, 
R.\,Fabbri, 
G.\,Falley, 
K.\,Gadow, 
E.\,Garutti,
P.\,G\"{o}ttlicher, 
S.\,Karstensen, 
F.\ Krivan,
K.\,Kschioneck, 
A.\,-I.\,Lucaci-Timoce,
B.\,Lutz, 
N.\,Meyer,
S.\,Morozov, 
V.\,Morgunov, 
M.\,Reinecke, 
F.\,Sefkow, 
P.\,Smirnov,
A.\,Vargas-Trevino, 
N.\,Wattimena, 
O.\,Wendt
\\ \it
DESY, Notkestrasse 85,
D-22603 Hamburg, Germany
}

\author{\centering  
N.\,Feege, 
J.\,Haller, 
S.\,Richter, 
J.\,Samson
\\ \it
Univ. Hamburg,
Physics Department,
Institut f\"ur Experimentalphysik,
Luruper Chaussee 149,
22761 Hamburg, Germany
}

\author{\centering 
A.\,Kaplan, H.\,-Ch.\,Schultz-Coulon, W.\,Shen, R.\,Stamen, A.\,Tadday
\\ \it
 University of Heidelberg, Fakultat fur Physik und Astronomie,
Albert Uberle Str. 3-5 2.OG Ost,
D-69120 Heidelberg, Germany
}

\author{\centering 
B.\,Bilki, E.\,Norbeck, Y.\,Onel
\\ \it
University of Iowa, Dept. of Physics and Astronomy,
203 Van Allen Hall, Iowa City, IA 52242-1479, USA
}

\author{\centering 
E.\,J.\,Kim
\\ \it
Chonbuk National University, Jeonju, 561-756, South Korea
}

\author{\centering 
G.\,Kim, D-W.\,Kim, K.\,Lee, S.\,C.\,Lee
\\ \it
Kangnung National University, HEP/PD, Kangnung, South Korea
}

\author{\centering 
B.\,van\,Doren,
G.\,W.\,Wilson
\\ \it
University of Kansas, Department of Physics and Astronomy,
Malott Hall, 1251 Wescoe Hall Drive, Lawrence, KS 66045-7582, USA
}

\author{\centering 
K.\,Kawagoe 
\\ \it
 Department of Physics, Kobe University, Kobe, 657-8501, Japan
}

\author{\centering 
P.\,D.\,Dauncey, 
\\ \it
Imperial College, Blackett Laboratory,
Department of Physics,
Prince Consort Road,
London SW7 2AZ, UK 
}

\author{\centering 
V.\,Bartsch\footnote{Now at University of Sussex, Physics and Astronomy Department, Brighton, Sussex, BN1 9QH, UK}, 
M.\,Postranecky, M.\,Warren, M.\,Wing
\\ \it
Department of Physics and Astronomy, University College London,
Gower Street,
London WC1E 6BT, UK
}

\author{\centering 
V.\, Boisvert,  
B.\,Green, 
A.\,Misiejuk, 
F.\,Salvatore\footnote{Now at University of Sussex, Physics and Astronomy Department, Brighton, Sussex, BN1 9QH, UK}
\\ \it
Royal Holloway University of London,
Dept. of Physics,
Egham, Surrey TW20 0EX, UK
}

\author{\centering 
E.\,Cortina Gil,
S.\,Mannai,
G.\,Nuessle 
\\ \it
Center for Particle Physics and Phenomenology (CP3)
Universit\'{e} catholique de Louvain, Belgium
}

\author{\centering 
M.\,Bedjidian   
A.\,Bonnevaux, 
C.\,Combaret, 
J.\,Fay,  
G.\,Grenier, 
J.C.\,Ianigro,
R.\,Kieffer, 
N.\,Lumb, 
I.\,Laktineh,  
P.\,Lebrun,  
H.\,Mathez,
M.\,Vander\,Donckt,
S.\,Vanzetto
\\ \it
Universit\'{e} de Lyon, F-69622, Lyon, France ;
Universit\'{e} de Lyon 1, Villeurbanne ;
CNRS/IN2P3, Institut de Physique Nucl\'{e}aire de Lyon
}

\author{\centering 
E.\,Calvo~Alamillo, M.\,C Fouz, J.\,Puerta-Pelayo 
\\ \it
CIEMAT, Centro de Investigaciones Energeticas, Medioambientales y Tecnologicas, Madrid. Spain 
}

\author{\centering 
D.\,S.\,Bailey, 
R.\,J.\,Barlow, 
R.\,J.\,Thompson 
\\ \it
The University of Manchester, School of Physics and Astronomy,
Schuster Lab,
Manchester M13 9PL,
UK
}

\author{\centering
M.\,Batouritski, 
O.\,Dvornikov, 
Yu.\,Shulhevich, 
N.\,Shumeiko, 
A.\,Solin,
P.\,Starovoitov, 
V.\,Tchekhovski, 
A.\,Terletski
\\ \it
National Centre of Particle and High Energy Physics of the
Belarusian State University, M.Bogdanovich str. 153, 220040 Minsk, Belarus
}

\author{\centering 
F.\,Corriveau 
\\ \it
Department of Physics, McGill University,
Ernest Rutherford Physics Bldg.,
3600 University Ave.,
Montr\'{e}al, Quebec,
CANADA H3A 2T8
}

\author{\centering 
V.\,Balagura, B.\,Bobchenko, 
M.\,Danilov, 
R.\,Mizuk, E.\,Novikov, V.\,Rusinov, E.\,Tarkovsky 
\\ \it
Institute of Theoretical and Experimental Physics, B. Cheremushkinskaya ul. 25,
RU-117218 Moscow, Russia
}

\author{\centering 
V.\,Andreev, N.\,Kirikova,  A.\,Komar, V.\,Kozlov, M.\,Negodaev, P.\,Smirnov, Y.\,Soloviev, A.\,Terkulov 
\\ \it
P.\,N.\, Lebedev Physical Institute,
Russian Academy of Sciences,
117924 GSP-1 Moscow, B-333, Russia
}

\author{\centering 
P.\,Buzhan, B.\,Dolgoshein, A.\,Ilyin, V.\,Kantserov, V.\,Kaplin, A.\,Karakash, E.\,Popova, S.\,Smirnov 
\\ \it
Moscow Physical Engineering Inst., MEPhI,
Dept. of Physics,
31, Kashirskoye shosse,
115409 Moscow, Russia
}

\author{\centering 
N.\,Baranova,
E.\,Boos, 
L.\,Gladilin,
D.\,Karmanov, 
M.\,Korolev, 
M.\,Merkin,
A.\,Savin,
A.\,Voronin
\\ \it
M.V.Lomonosov Moscow State University, D.V.Skobeltsyn Institute of Nuclear
Physics (SINP MSU),
1/2 Leninskiye Gory, Moscow, 119991, Russia
}

\author{\centering 
A.\,Topkar
\\ \it
Bhabha Atomic Research Center,
Mumbai 400085, India
}

\author{\centering 
C.\,Kiesling,
S.\,Lu, 
O.\,Reimann, 
K.\,Seidel, 
F.\,Simon
C.\,Soldner, 
L.\,Weuste
\\ \it
Max Planck Inst. f\"ur Physik,
F\"ohringer Ring 6,
D-80805 Munich, Germany
}

\author{\centering 
J.\,Bonis, 
B.\,Bouquet,    
S.\,Callier, 
P.\,Cornebise, 
Ph.\,Doublet,
F.\,Dulucq, 
M.\,Faucci Giannelli, 
J.\,Fleury,
G.\,Guilhem, 
H.\,Li,  
G.\,Martin-Chassard, 
F.\,Richard, 
Ch.\,de la Taille, 
R.\,Poeschl, 
L.\,Raux,  
N.\,Seguin-Moreau, 
F.\,Wicek, 
Z.\,Zhang
\\ \it
Laboratoire de L'acc\'elerateur Lin\'eaire,
Centre d'Orsay, Universit\'e de Paris-Sud XI,
BP 34, B\^atiment 200,
F-91898 Orsay CEDEX, France
}

\author{\centering 
M.\,Anduze, 
K.\,Belkadhi,
V.\,Boudry, 
J-C.\,Brient, 
C.\,Clerc, 
R.\,Cornat,
D.\,Decotigny,
F.\,Gastaldi, 
D.\,Jeans, 
A.\,Karar,  
P.\,Mora de Freitas, 
G.\,Musat, 
M.\,Reinhard, 
A.\,Roug\'{e},
M.\,Ruan,  
J-Ch.\,Vanel, 
H.\,Videau
\\ \it
      Laboratoire Leprince-Ringuet (LLR)  -- \'{E}cole Polytechnique,
      CNRS/IN2P3,
      Palaiseau, F-91128 France
}


\author{\centering 
K-H.\,Park
\\ \it
Pohang Accelerator Laboratory, Pohang 790-784, South Korea
}

\author{\centering 
J.\,Zacek 
\\ \it
Charles University, Institute of Particle \& Nuclear Physics,
V Holesovickach 2,
CZ-18000 Prague 8, Czech Republic  
}

\author{\centering 
J.\,Cvach, 
P.\,Gallus, 
M.\,Havranek, 
M.\,Janata, 
J.\,Kvasnicka,
M.\,Marcisovsky, 
I.\,Polak, 
J.\,Popule, 
L.\,Tomasek, 
M.\,Tomasek, 
P.\,Ruzicka, 
P.\,Sicho, 
J.\,Smolik, 
V.\,Vrba, 
J.\,Zalesak 
\\ \it
Institute of Physics, Academy of Sciences of the Czech Republic, Na Slovance 2,
CZ-18221 Prague 8, Czech Republic
}

\author{\centering 
Yu.\,Arestov
V.\,Ammosov, B.\,Chuiko, V.\,Gapienko,
Y.\,Gilitski,V.\,Koreshev, A.\,Semak, Yu.\,Sviridov, V.\,Zaets
\\ \it
Institute of High Energy Physics,
Moscow Region,
RU-142284 Protvino,
Russia
}

\author{\centering 
B.\,Belhorma, M.\, Belmir, H.\,Ghazlane
\\ \it
Centre National de l'Energie, des Sciences et des Techniques Nucl\'{e}aires, 
B.P. 1382, R.P. 10001, Rabat, Morocco
}

\author{\centering 

R.\,Coath, J.\,P.\,Crooks,M.\,Stanitzki,  J.\,Strube, R.\,Turchetta, M.\,Tyndel, Z.\,Zhang
\\ \it
Rutherford Appleton Laboratory, Chilton, Didcot,
Oxon OX110QX, UK 
}

\author{\centering 
M.\,Barbi  
\\ \it
University of Regina, 
Department of Physics,
Regina, Saskatchewan,
Canada S4S 0A2
}

\author{\centering 
S.\,W.\,Nam, I.\,H.\,Park, J.\,Yang 
\\ \it
Ewha Womans University, Dept. of Physics,
Seoul 120,
South Korea
}

\author{\centering 
Jong-Seo Chai, Jong-Tae Kim, Geun-Bum Kim
\\ \it
Sungkyunkwan University,
300 Cheoncheon-dong, Jangan-gu, Suwon, Gyeonggi-do  440-746, South Korea
}

\author{\centering 

Y.\,Kim
\\ \it
Korea Institute of Radiological and
Medical Sciences,
215-4 Gangeung-dong,
Nowon-gu, Seoul 139-706,
SOUTH KOREA
}

\author{\centering 
J.\,Kang, Y.\,-J.\,Kwon  
\\ \it
Yonsei  University, Dept. of Physics,
134 Sinchon-dong,
Sudaemoon-gu, Seoul 120-749,
South Korea
}

\author{\centering 
Ilgoo Kim, Taeyun Lee, Jaehong Park, Jinho Sung
\\ \it
School of Electric Engineering and Computing Science, Seoul National University,
Seoul 151-742, South Korea
}

\author{\centering              
S.\, Itoh,  K.\,Kotera, M.\, Nishiyama, T.\,Takeshita
\\ \it
Shinshu Univ.\,,
Dept. of Physics,
3-1-1 Asaki,
Matsumoto-shi, Nagano 390-861,
Japan
}

\author{\centering 
A.\,Khan, D.\,H.\,Kim, J.E.\,Kim, D.\,J.\,Kong, Y.D.\,Oh, S.\,Uozumi
\\ \it
Kyungpook National Univ., Dept. of Physics, 1370 San Kyuk-dong, Puk ku, Taegu 635, SOUTH KOREA
}

\author{\centering              
H.\,Koike, 
Y.\,Sudo, 
Y.\,Takahashi, 
K.\,Tanaka, 
F.\,Ukegawa
\\ \it
  University of Tsukuba, Graduate School of Pure and Applied Sciences,
  Tennoudai 1-1-1, Tsukuba, Ibaraki 305-8571, Japan
}

\author{\centering 
S.\,Weber, C.\,Zeitnitz
\\ \it
Bergische Universit\"{a}t Wuppertal
Fachbereich 8 Physik,
Gaussstrasse 20,
D-42097 Wuppertal, GERMANY
}






\end{document}